\documentclass[twocolumn]{aastex63}
\usepackage{lineno}
\usepackage{amsmath}
\usepackage{hyperref}
\usepackage{soul}

\begin{document}

\title{Detection of diffuse { gamma-ray} emission towards a massive star forming region hosting Wolf-Rayet stars}

\author[0000-0002-5582-8265]{Kai Wang}
\affil{School of Astronomy and Space Science, Nanjing University, Nanjing 210023, Jiangsu, China; xywang@nju.edu.cn,hmzhang@nju.edu.cn}
\affil{Key laboratory of Modern Astronomy and Astrophysics(Nanjing University), Ministry of Education, Nanjing 210023, People Republic of China}

\author[0000-0001-6863-5369]{Hai-Ming Zhang}
\affil{School of Astronomy and Space Science, Nanjing University, Nanjing 210023, Jiangsu, China; xywang@nju.edu.cn,hmzhang@nju.edu.cn}
\affil{Key laboratory of Modern Astronomy and Astrophysics(Nanjing University), Ministry of Education, Nanjing 210023, People Republic of China}

\author[0000-0003-1576-0961]{Ruo-Yu Liu}
\affil{School of Astronomy and Space Science, Nanjing University, Nanjing 210023, Jiangsu, China; xywang@nju.edu.cn,hmzhang@nju.edu.cn}
\affil{Key laboratory of Modern Astronomy and Astrophysics(Nanjing University), Ministry of Education, Nanjing 210023, People Republic of China}
\author[0000-0002-5881-335X]{Xiang-Yu Wang}
\affil{School of Astronomy and Space Science, Nanjing University, Nanjing 210023, Jiangsu, China; xywang@nju.edu.cn,hmzhang@nju.edu.cn}
\affil{Key laboratory of Modern Astronomy and Astrophysics(Nanjing University), Ministry of Education, Nanjing 210023, People Republic of China}

\begin{abstract}
{ Isotopic and elemental abundances seen in Galactic cosmic rays imply} that $\sim20\%$ of the cosmic-ray (CR) nuclei are
probably synthesized by massive Wolf-Rayet (WR) stars. Massive star clusters hosting WR and OB-type stars have been proposed as potential Galactic cosmic-ray accelerators for decades, { in particular via diffusive shock acceleration  at wind termination shocks}. Here we report the analysis of {\em Fermi} Large Area Telescope's data towards the direction  of  Masgomas-6a, a young massive star cluster candidate  hosting two WR stars. { We detect an extended $\gamma$-ray source with $\rm{TS}=183$ in the vicinity of Masgomas-6a}, spatially coincident with two unassociated {\em Fermi} 4FGL sources.
We also  present the CO observational results of molecular clouds in this region, using the data from the Milky Way Imaging Scroll Painting project. The { $\gamma$-ray} emission intensity correlates well with the distribution of molecular gas at the distance of Masgomas-6a, { indicating that these gamma rays may be produced by CRs accelerated by massive stars in Masgomas-6a}. At the distance of $3.9{\rm \ kpc}$ of Masgomas-6a, {the luminosity of the extended source { is $(1.81\pm0.02)\times 10^{35}{\rm \ erg \ s^{-1}}$}}. With a kinetic luminosity of $\sim 10^{37}{\rm erg \ s^{-1}}$ in the stellar winds,  the  WR stars are capable of powering the { $\gamma$-ray} emission { via neutral pion decay resulted from cosmic ray $pp$ interactions}. The size of the GeV source and the energetic requirement suggests a CR diffusion coefficient smaller than that in the Galactic { interstellar medium}, indicating strong suppression of CR diffusion in the  molecular cloud.

\end{abstract}


\section{Introduction}           
\label{sect:intro}

The { elemental composition} of Galactic cosmic rays (GCRs) has been investigated with several experiments (IMP7 \citep{1979ApJ...232L..95G}, ISEE-3 \citep{1981PhRvL..46..682W}, Voyager \citep{1994ApJ...426..366L}, ACE-CRIS \citep{2005ApJ...634..351B}, { DAMPE \citep{2019SciA....5.3793A}, AMS \citep{2020PhRvL.124u1102A}} and etc.). For most elements, the isotopic ratios seen in GCRs are similar to
those in the solar wind { \citep{1979ICRC....1..412W}}. The most notable exception to this is $^{22}$Ne/$^{20}$Ne, where the CR value
is about 5 times that of the solar wind {\citep{2005ApJ...634..351B,2008NewAR..52..427B}}. According to the most recent models of nucleosynthesis, a
large amount of $^{22}$Ne is generated in Wolf-Rayet (WR) stars {\citep{2003ApJ...590..822H,2019JPhCS1400b2011K}}. The interacting winds of massive stars have been recognized as potential CR accelerators as early as in the 1980s {{\citep{1982IzSSR..46.1659B,1983SSRv...36..173C}}}.
The acceleration could take place in the vicinity of the stars \citep{2014A&ARv..22...77B, 2019NatAs...3..561A}, { in the colliding wind region of WR and OB binary systems \citep{2003A&A...399.1121B}}, or in superbubbles, multi-parsec
structures caused by the collective activity of massive stars \citep{2004A&A...424..747P,2014A&ARv..22...77B}.
\citet{2003ApJ...590..822H} found that the observed value of $^{22}$Ne/$^{20}$Ne can be achieved with a
mixture of $\sim20\%$ WR star material and $\sim80\%$ material with standard composition. Using
state-of-the-art stellar evolution models,  \citet{2019JPhCS1400b2011K} calculated the amount of neon isotopes produced by
 interacting winds of OB-type and WR stars and also concluded that massive star clusters can produce a significant fraction of GCRs
and $^{22}$Ne.

Spectroscopic data show that WR stars, along with OB-type stars, have powerful stellar
winds with velocities of 1000-3000 km/s { \citep{2019A&A...625A..57H}}. Due to their small size and high star density \citep{2010ARA&A..48..431P},
young massive star clusters are likely to have strong colliding magnetohydrodynamic shock flows,
where efficient particle acceleration can take place. Recently, \citet{2018JKAS...51...37S} estimated the total
power of OB-type and WR stars’ winds in the Galaxy as $\sim 1.1\times 10^{41}{\rm erg \ s^{-1}}$. If $(1-10)\%$ of the wind
luminosity is converted to GCRs, these stellar winds can provide a significant contribution
to the GCR production below the knee.
The conversion of stellar wind power to CRs might be traced by secondary $\gamma$-rays, the products of interactions
of CRs with the surrounding gas. The diffuse GeV $\gamma$-rays detected by \emph{Fermi}-LAT telescope around
some compact clusters, such as Cygnus OB2\citep{2011Sci...334.1103A}, NGC 3603 \citep{2017A&A...600A.107Y}, Westerlund 2 \citep{2018A&A...611A..77Y} and RSGC 1 \citep{Sun2020}, can be naturally interpreted within this scenario.
{ Detection of TeV or higher energy photons from some massive star clusters (e.g., the Cygus region) may suggest that cosmic rays are accelerated up to energy of 100 TeV to PeV \citep{2002A&A...393L..37A,2021NatAs...5..465A,2021PhRvL.127c1102A}}

Recently, \citet{2018A&A...614A.116R} identified two groups of massive stars aligned in the $l \sim 38^{\circ}$ Galactic direction. They find more than 20  massive stars, including two Wolf-Rayets (WR122-11 and the new WR122-16), OB-type stars and one transitional object (the Luminous Blue Variable (LBV) candidate IRAS 18576+0341 \citep{2002A&A...382.1005P}). The individual distances and radial velocities of these massive stars indicate two populations of massive stars in the same line of sight: Masgomas-6a and Masgomas-6b. {Masgomas-6a is reported as a massive star cluster candidate with a total mass (lower limit) of  $(9.0-1.3) \times 10^{3} M_{\odot}$, located at $3.9^{+0.4}_{-0.3}$ kpc, contains both Wolf-Rayets and several OB-type stars}. Masgomas-6b, located at $9.6\pm0.4$ kpc,  has a total mass (lower limit) of $(10.5-1.5) \times 10^{3} M_{\odot}$ and hosts a LBV candidate and an evolved population of supergiants.  For both objects, the presence of evolved massive stars (the WN8-9h Wolf-Rayet in
Masgomas-6a and the LBV candidate in Masgomas-6b) sets an age limit of 5 Myr \citep{2018A&A...614A.116R}.

In this paper, we report the detection of an extended { $\gamma$-ray} source  towards the direction  of  Masgomas-6 with the \emph{Fermi} Large Area Telescope. We also  present the CO observational results of molecular clouds (MCs) in this region, using the data from the Milky Way Imaging Scroll Painting (MWISP) project, which is a multi-line survey in $^{12}$CO/$^{13}$CO/C$^{18}$O along the northern Galactic plane with the Purple Mountain Observatory (PMO) 13.7 m telescope. The correlation between the { $\gamma$-ray} intensity and the molecular gas at the distance of Masgomas-6a suggests that the GeV { $\gamma$-ray} source is likely related with the massive stars in Masgomas-6a.  The paper is organized as follows. We first present the {\em Fermi}/LAT data analysis result of the { $\gamma$-ray} source in \S 2. We then  present the MWISP observational results of molecular clouds in this region in \S 3. In \S 4, the interpretation of the GeV source and the implication is presented. Finally, we give a summary in \S 5.

\section{\emph{Fermi}/LAT Data Analysis}

The \textit{Fermi}-LAT is a wide field-of-view, high-energy { $\gamma$-ray} telescope, covering the energy range from below 20 MeV to more than 300 GeV, and it has continuously monitored the sky since 2008 \citep{Atwood2009}. The  Pass 8 data taken from 2008 August 4 to 2022 April 9 are used to study the GeV emission towards  Masgomas-6, where two {\em Fermi} sources (4FGL J1900.4+0339 and 4FGL J1858.0+0354) are found in the vicinity.
All { $\gamma$-ray} photons within a $14\degr \times14\degr$ region of interest (ROI) centered on the position of Masgomas-6 are considered for the binned maximum likelihood analysis.
The publicly available software \textit{Fermipy} (ver.1.0.1) and \textit{Fermitools} (ver.2.0.8) are used to perform the data analysis.
The event type FRONT + BACK  and the instrument response functions (IRFs) (\textit{$P8R3\_SOURCE\_V3$}) are used. 
We only consider the $\gamma$-ray events in the $3-500 \, \rm GeV$ energy range, with the standard data quality selection criteria ``$(DATA\_QUAL > 0)  \&\& (LAT\_CONFIG == 1)$". 
To minimize the contamination from { $\gamma$-rays} from the Earth limb, the maximum zenith angle is adopted to be 90$\degr$.
We include all sources listed in the fourth \textit{Fermi}-LAT catalog\citep{Abdollahi2020a} and the diffuse Galactic interstellar emission (IEM, $gll\_iem\_v07.fits$),  isotropic emission (``$iso\_P8R3\_SOURCE\_V3\_v1.txt$'' ) in the background model.
The spectral parameters of the sources within $4\degr$ of Masgomas-6, Galactic and isotropic diffuse emission components are all set free.
The significance of $\gamma$-ray sources are estimated by maximum likelihood test statistic (TS), which is defined by TS$= -2 (\ln\mathcal{L}_{0}-\ln\mathcal{L}_{1})$, where $\mathcal{L}_{1}$ and $\mathcal{L}_{0}$ are maximum likelihood values for the background with target source and without the target source (null hypothesis). The square root of the TS is approximately equal to the detection significance of a given source.

\subsection{Morphological analysis}
\begin{figure*}
\includegraphics[angle=0,scale=0.28]{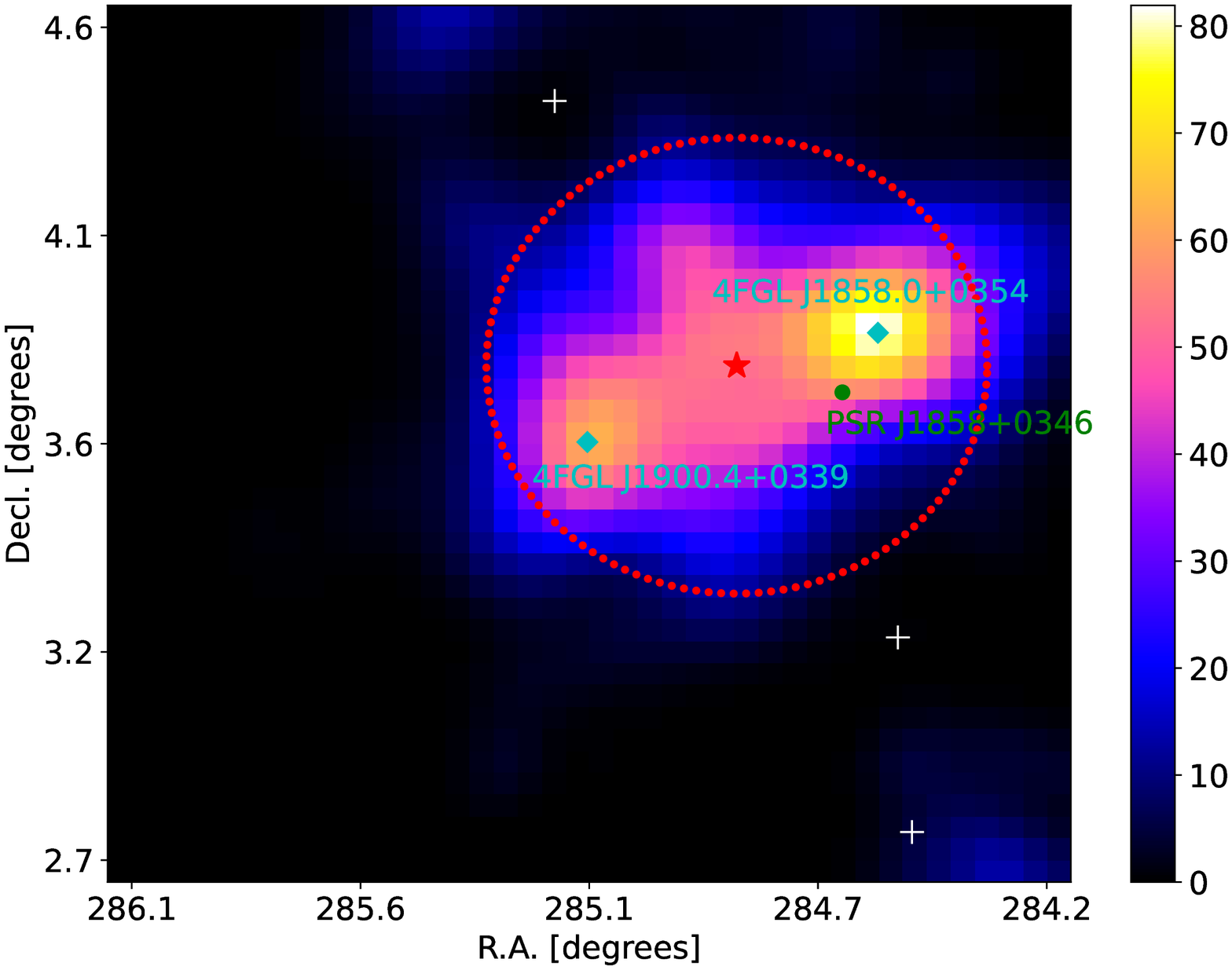}
\includegraphics[angle=0,scale=0.28]{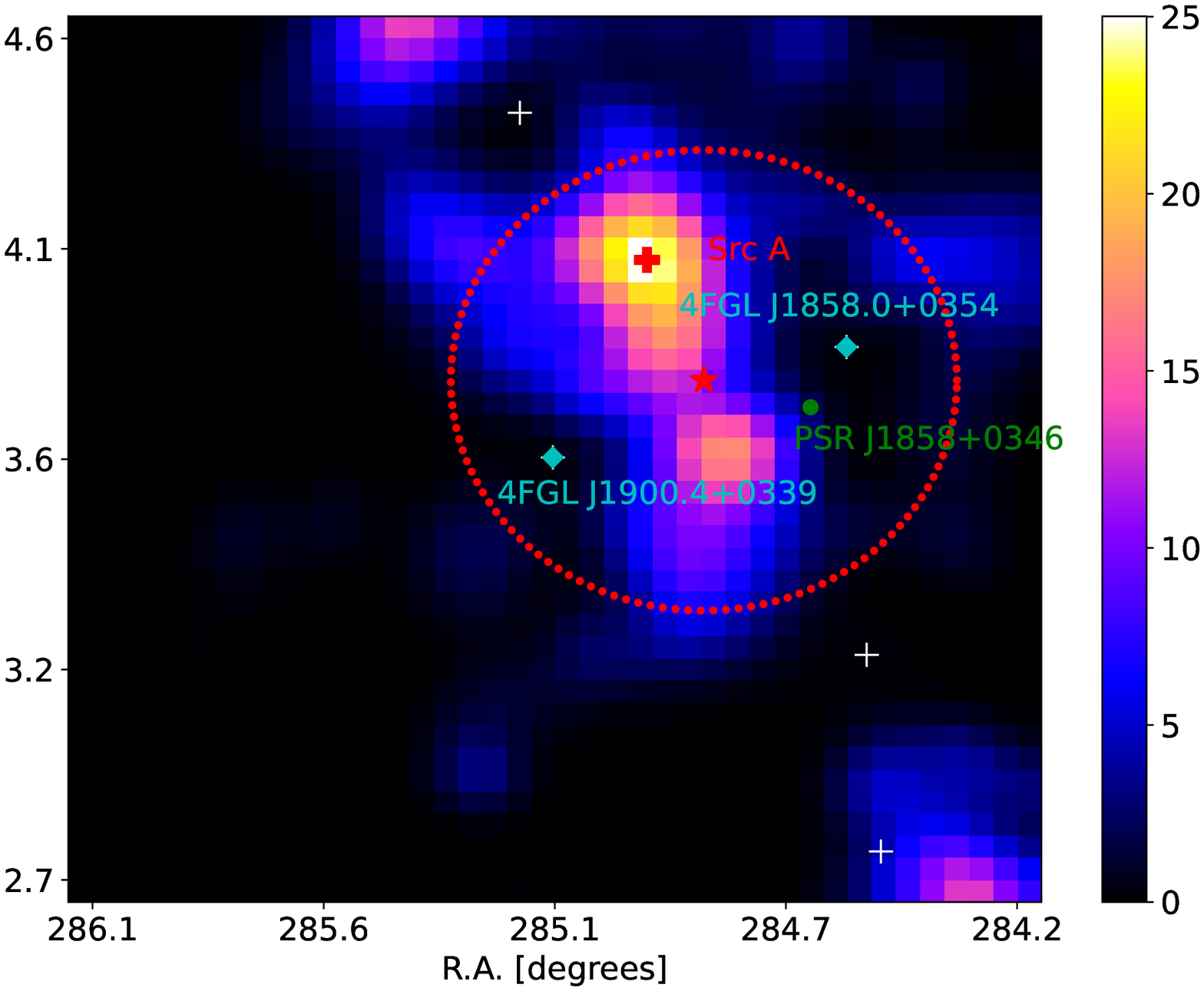}
\includegraphics[angle=0,scale=0.28]{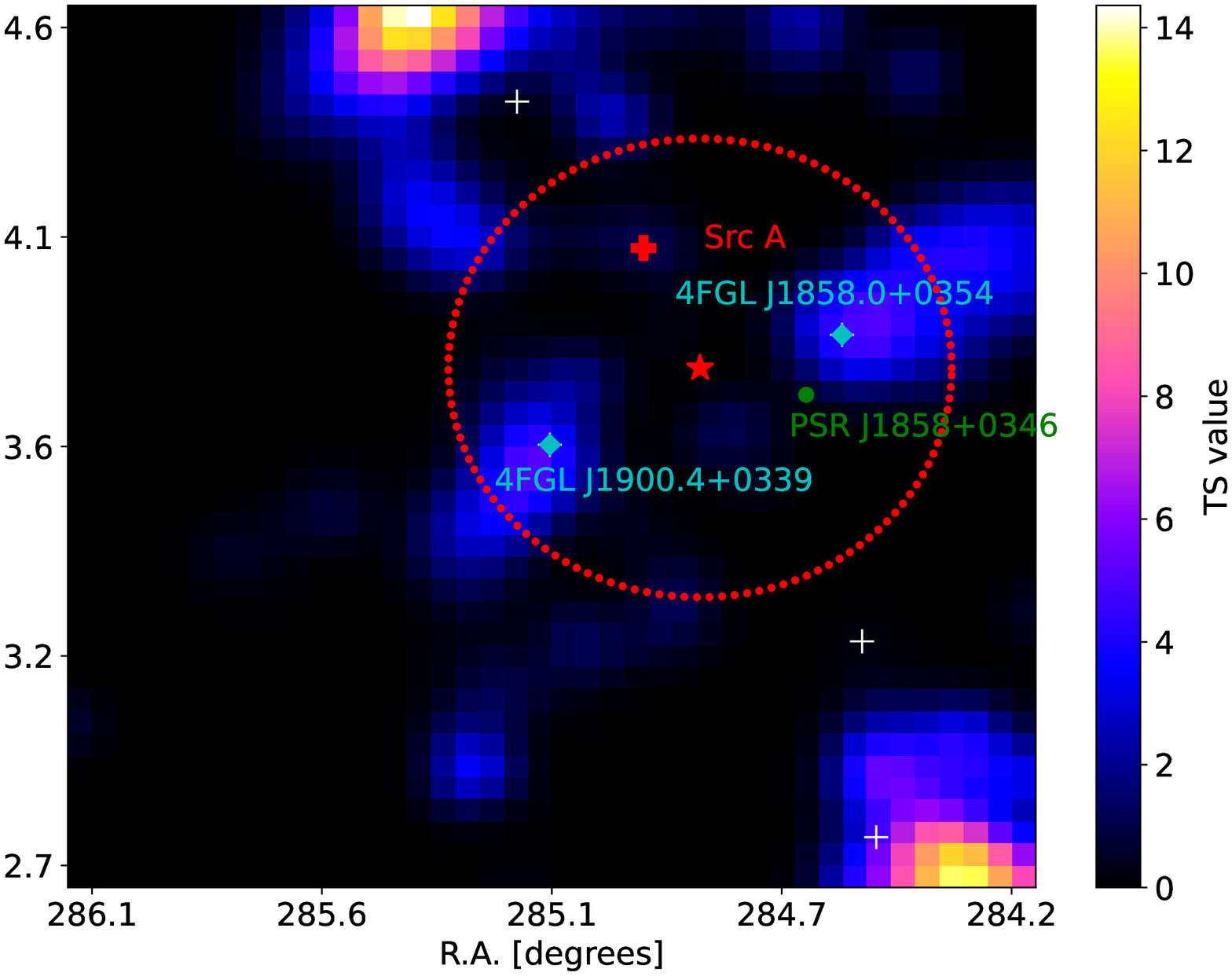}
\caption{Left panel: \emph{Fermi}-LAT TS map of the region around Masgomas-6a  in  3-500 GeV. Middle panel: the residual TS map of the two point-like source model. Right panel: the residual TS map of the disk model. Background 4FGL sources are shown with { white} crosses. The position of the PSR J1858+0346 is shown with a { green} dot. 
The red dashed circle shows the radius of the disk in the best-fit uniform disk model. 
}
\label{GeVmap0}
\end{figure*}

{ To obtain a sufficiently good angular resolution for the morphology analysis, we use the LAT data in  3-500 GeV.}
Figure \ref{GeVmap0} shows the 3-500 GeV TS map in the the vicinity of Masgomas-6 with the binned likelihood analysis. 
The two 4FGL point-like sources near Masgomas-6, 4FGL J1900.4+0339 and 4FGL J1858.0+0354, are located $0.16\degr$ and $0.52\degr$ away from the center of Masgomas-6, respectively.
{ We plan to investigate what appear to be diffuse emissions beyond the two 4FGL point-like sources.} Since these two point-like sources are unassociated with any known counterparts and there is  diffuse emission  in the vicinity, we investigate whether the $\gamma$-ray emission in this region represents a spatially extended source. 
We use the \emph{Fermipy} tool to quantitatively evaluate the extension and location of this $\gamma$-ray emission.
A uniform disk model and a two-dimensional (2D) Gaussian model are used to evaluate the extension. 
The results are shown in Table \ref{tab:morph}.
We find that the uniform disk model describes well the GeV morphology, with the best-fit extension of $0.43\degr_{-0.03\degr}^{+0.02\degr}$ and the best-fit position at ($\rm R.A.,Decl.$)=($284.81\degr\pm0.08\degr$,$3.84\degr\pm0.04\degr$) in the energy band above 3 GeV. This position is $0.22\degr$ away from the position of  Masgomas-6. 
The extension represents the radius containing 68\% of the intensity.
The $\rm TS_{ext}$ is estimated to be 129.80 for the disk model, which is defined as $\rm {TS_{ext}}=2(\ln\mathcal{L}_{ext}-\ln\mathcal{L}_{ps})$, where $\mathcal{L}_{ext}$ is the maximum likelihood value for the extend model and $\mathcal{L}_{ps}$ is the maximum likelihood value for the point-like model \citep{Ackermann2017}.
According to \citet{Ackermann2017}, the criterion to define a source as being extended is  $\rm TS_{ext}\geq16$ and  $\rm TS_{ext}\geq TS_{2pts}$ (here $\rm {TS_{2pts}}=2(\ln\mathcal{L}_{2pts}-\ln\mathcal{L}_{ps})$ represents the improvement when adding a second point source). 
We find that the { $\gamma$-ray} source, assuming a uniform disk model, meets the criterion  with $\rm TS_{ext}=129.80$ and $\rm TS_{2pts}= 106.62$ (see Table \ref{tab:morph}).
We also test the 2D Gaussian model and find that it is also better than the the point-like model. Therefore, the GeV emission in this region is described better by an extended source. 

Furthermore, we compare the residual TS maps after subtracting the disk component and the two point-like source component. We find that there is bright (maximum ${\rm TS} \sim 25$) emission (referred to as Src A, see Figure \ref{GeVmap0}) at the top left corner of the residual TS map in the two point-like source model, { but the significance of this emission is significantly reduced } in the disk model. This implies that the extended source model fits the GeV emission better than the two point-like sources model, consistent with our above analysis.

\subsection{Energy Spectrum}

After the morphology was fixed, the extended disk model is used to study the spectrum of the source. The  spectral energy distribution (SED) of the  source in the energy band $>$ 0.1 GeV, shown in Figure \ref{sed}, is derived by \emph{gtlike} assuming the best-fit uniform disk extension.
When the TS value of spectra data point is less than 4, an upper limit is calculated at 95\% confidence level using a Bayesian method \citep{Helene1983}. Compared to a single power-law, the spectrum is better described by a log parabola empirical function $dN/dE = N_0(E/E_{b})^{-(\alpha+\beta \text{log}(E/E_{b}))}$. 
The spectral index and break energy are found to be $\alpha=2.29\pm0.01$, $\beta=0.19\pm0.01$ and $E_{b}=1268.47\pm2.81 \rm {MeV}$. The energy flux is $\rm (9.92\pm0.12)\times 10^{-11} \ erg\ cm^{-2} \ s^{-1}$ in 0.1-500 GeV. Assuming a distance of $d=3.9{\rm \ kpc}$, the total { $\gamma$-ray} luminosity of the source in 0.1-500 GeV is $\rm (1.81\pm0.02)\times 10^{35} \ erg\ s^{-1}$.  The preference of a curved spectrum over a single power-law is supported by  $\rm TS_{curve}=102.75$ ($10.14 \sigma$), where $\rm TS_{curve}$ is defined as $\rm{TS_{curve}}=2(\ln\mathcal{L}_{curved \  spectrum}-\ln\mathcal{L}_{power-law})$.
In the energy band greater than 3 GeV, however, the spectra of the disk model is well described by a single power-law function with $\rm TS_{curve}<9$ ($\rm TS_{curve}=9$ corresponding to $3\sigma$ \citep{Abdollahi2020a}).

\section{CO observations}

We make use of the data from the Milky Way Imaging Scroll Painting (MWISP\footnote{\url{http://english.dlh.pmo.cas.cn/ic/}}) project, which is a multi-line survey in $^{12}$CO/$^{13}$CO/C$^{18}$O observed simultaneously using the 13.7 m millimeter-wavelength telescope of the Purple Mountain Observatory at Delingha. 
The detailed observing strategy, the instrument, the spectral resolution, and the quality of the CO observations can be found in \citet{Su2019}. 
In this work, we present the results of the MWISP CO survey for $2\degr \times 2\degr$ regions centering at $(l,b)=(37.27\degr,-0.23\degr)$.
We estimated the kinematic distance to the MCs using the Milky Way's rotation curve suggested by \citet{Sofue2015}, assuming the Sun's Galactocentric distance to be 8.0 kpc and  Solar rotation velocity to be $238 \ \rm km \ s^{-1}$.

\begin{figure*}
\includegraphics[angle=0,scale=0.4]{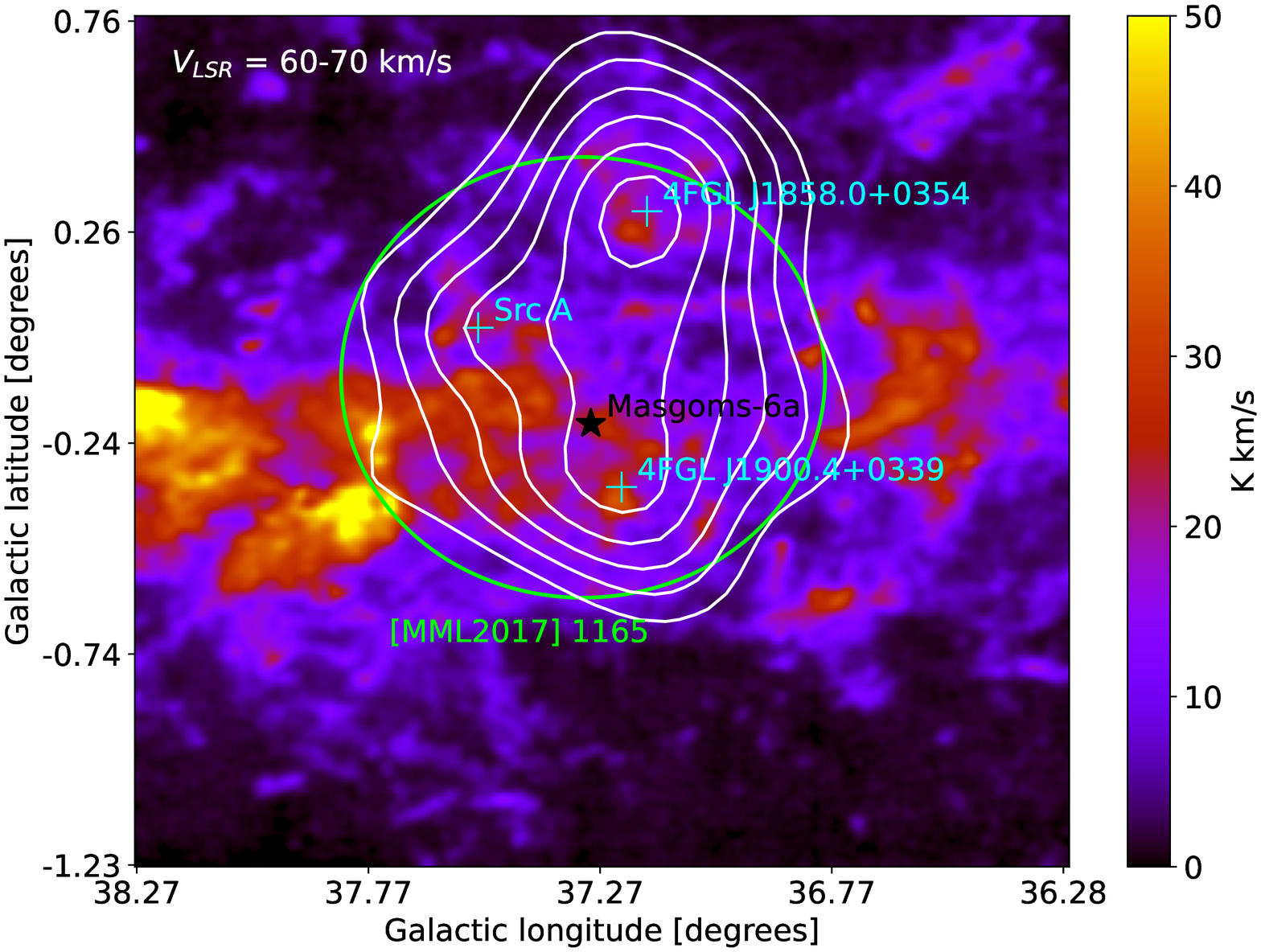}
\includegraphics[angle=0,scale=0.4]{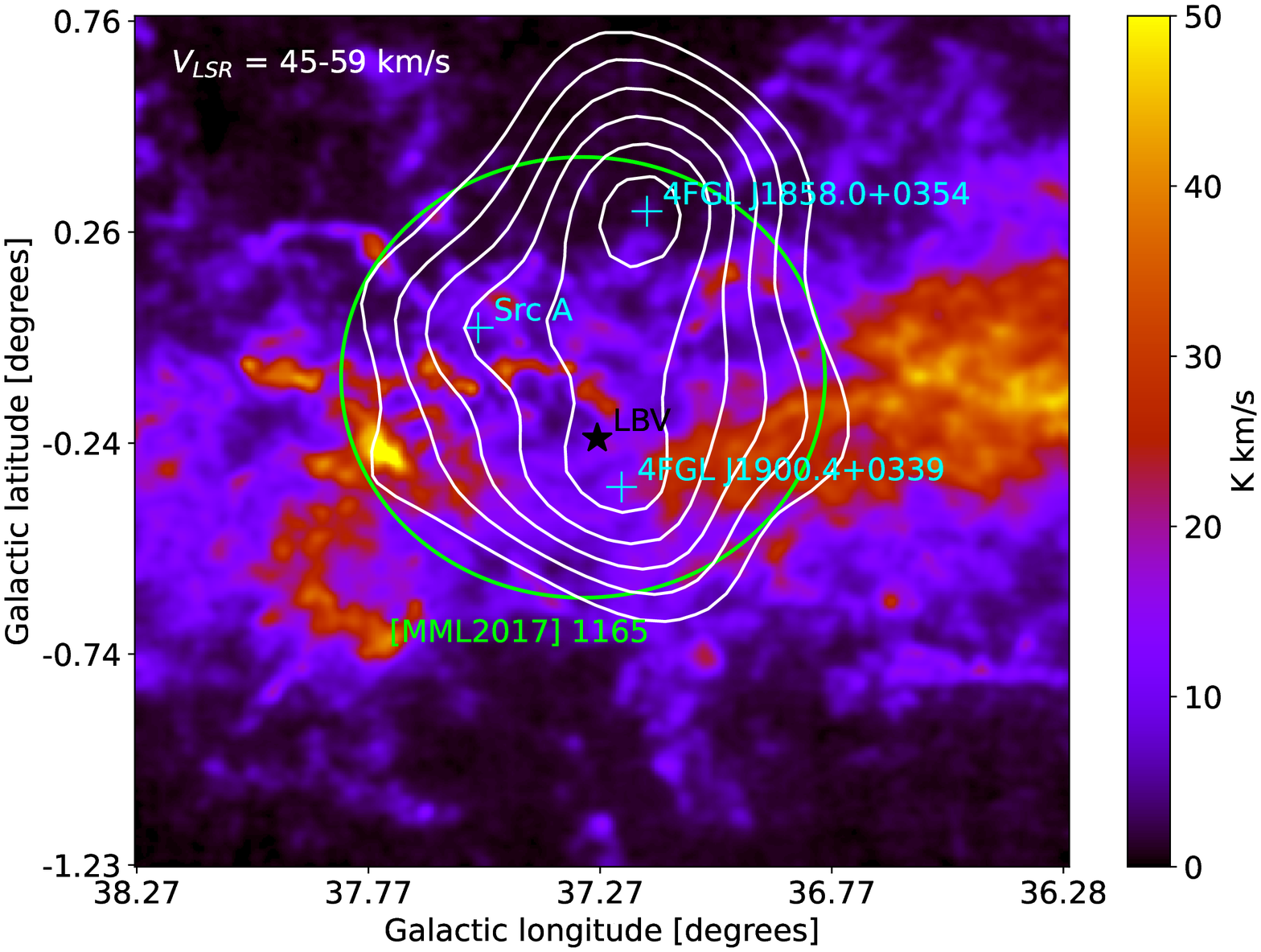}
\caption{$^{12}$CO($J$=1--0) intensity map integrated in velocity range $60-70 \ \rm km \ s^{-1}$ (left panel) and $45-59 \ \rm km \ s^{-1}$ (right panel). White contours correspond to {\em Fermi}-LAT significance map starting from $3\sigma$ to $8\sigma$ by $1\sigma$ steps. { Crosses in cyan indicate the location of point sources and Src A.} The { green circle} shows the radius of the molecular cloud "[MML2017] 1165" named by \citet{Miv2017}. "[MML2017] 1165" is located in the velocity range of $\rm V_{LSR}= (63.52\pm6.41) \, km\, s^{-1}$.
}
\label{COmapall}
\end{figure*}

The distance of Masgomas-6a is estimated to be $3.9^{+0.4}_{-0.3}$ kpc \citep{2018A&A...614A.116R},  corresponding to a velocity $V_{\rm LSR}$ in the  range of  $\sim60-70 \ \rm km \ s^{-1}$ for the MC CO emission. Note that, the velocity of each MC could indicates two candidate kinematic distances, the near side one and the far side one. The far distance of the MC in $\sim60-70 \ \rm km \ s^{-1}$ is 8.4--9.0 kpc. The distance of Masgomas-6b is estimated to be $9.6\pm0.4 \  \rm kpc$, corresponding to a velocity $V_{\rm LSR}$ in the range of $\sim 45-59 \rm \ km \ s^{-1}$.
The $^{12}$CO(J = 1-0) maps for the velocity range 60 to 70 $\rm km \ s^{-1}$ and 45 to 59 $\rm km \ s^{-1}$ are, respectively, shown in the left and right panel of Figure \ref{COmapall}.
{ The intensity maps of the gamma-ray emission correlate more strongly with the gas distribution in the velocity range of 60 to 70 $\rm km \ s^{-1}$}. Particularly, the two point-like 4FGL  sources (4FGL J1900.4+0339 and 4FGL J1858.0+0354) and Src A each coincide  with one of the densest regions of the gas distribution. On the other hand, { $\gamma$-ray emission intensity correlates poorly with gas distribution in the 45-59 $\rm km \ s^{-1}$ velocity range compared to that in the 60-70 $\rm km \ s^{-1}$ velocity range.} This indicates that the extended { $\gamma$-ray} source is more likely to be related to Masgomas-6a. 
\citet{Miv2017} produced a catalog of 8107 molecular clouds that covers the entire Galactic plane and includes 98\% of the $^{12}$CO emission observed within $b\pm5\degr$, using a hierarchical cluster identification method applied to the result of a Gaussian decomposition of CO data. 
We note that a molecular cloud in this catalog (named ``[MML2017] 1165'') \footnote{\url{http://simbad.u-strasbg.fr/simbad/sim-ref?querymethod=bib&simbo=on&submit=submit+bibcode&bibcode=2017ApJ...834...57M}} is located in the velocity range of $\rm V_{LSR}= (63.52\pm6.41) \, km\, s^{-1}$, consistent with our measurement. The angular radius of this cloud is $0.518\degr$ (see the magenta  circle in the left panel of Figure \ref{COmapall}). Interestingly, the position and size of this MC agree well with those of the extended GeV source, supporting the association between the GeV source and the MC.

\begin{figure*}
\includegraphics[angle=0,scale=0.51]{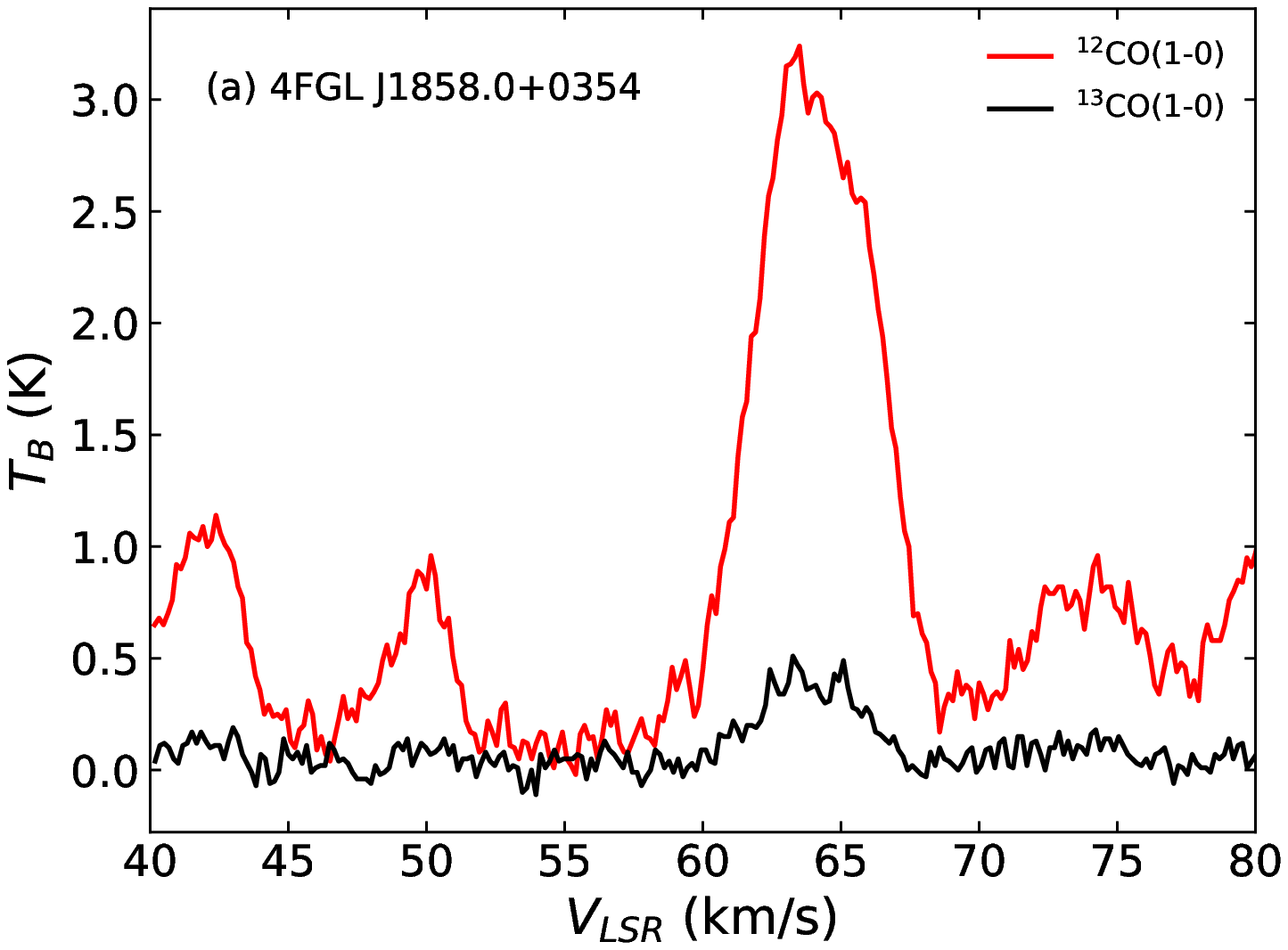}
\includegraphics[angle=0,scale=0.51]{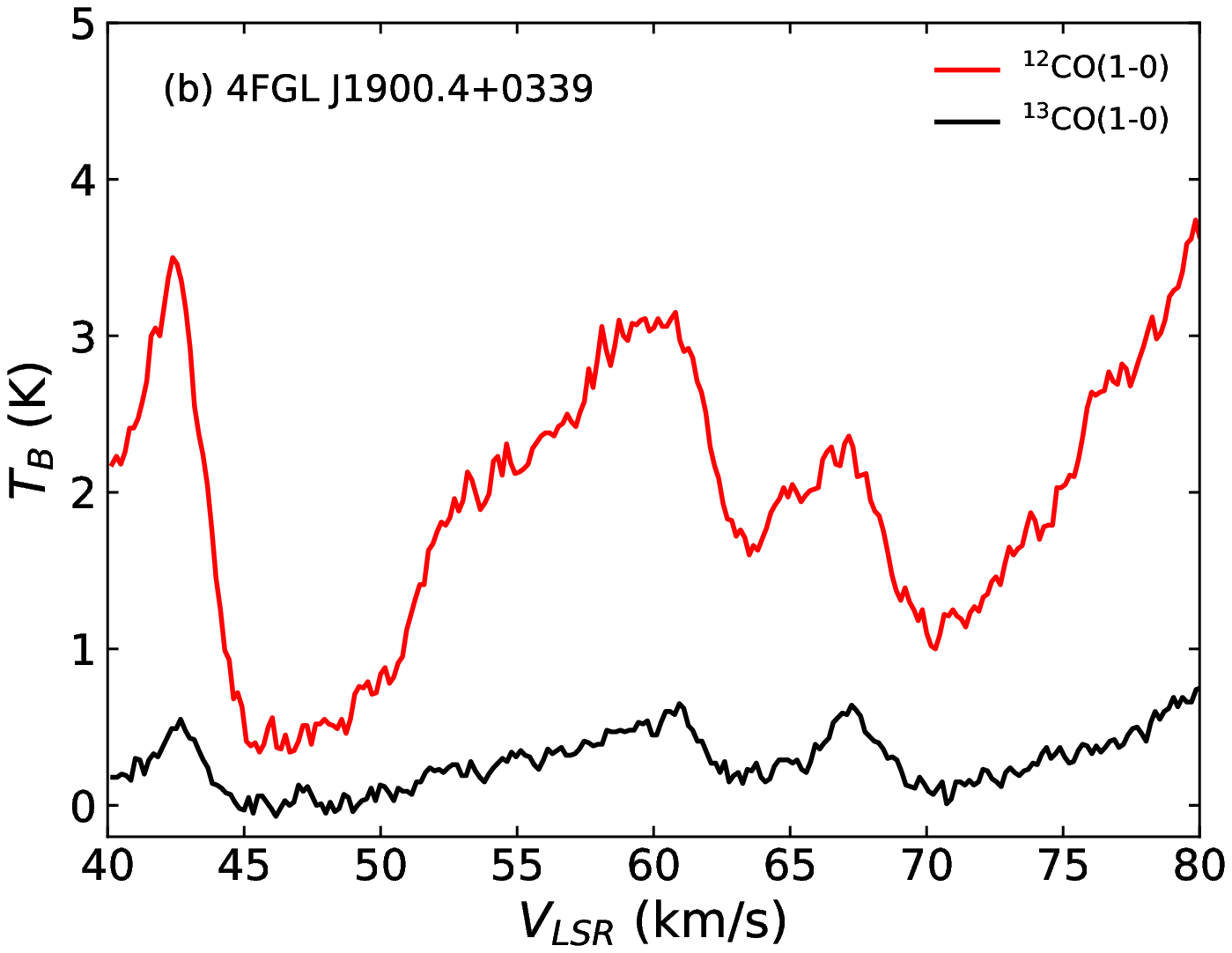}\\
\includegraphics[angle=0,scale=0.51]{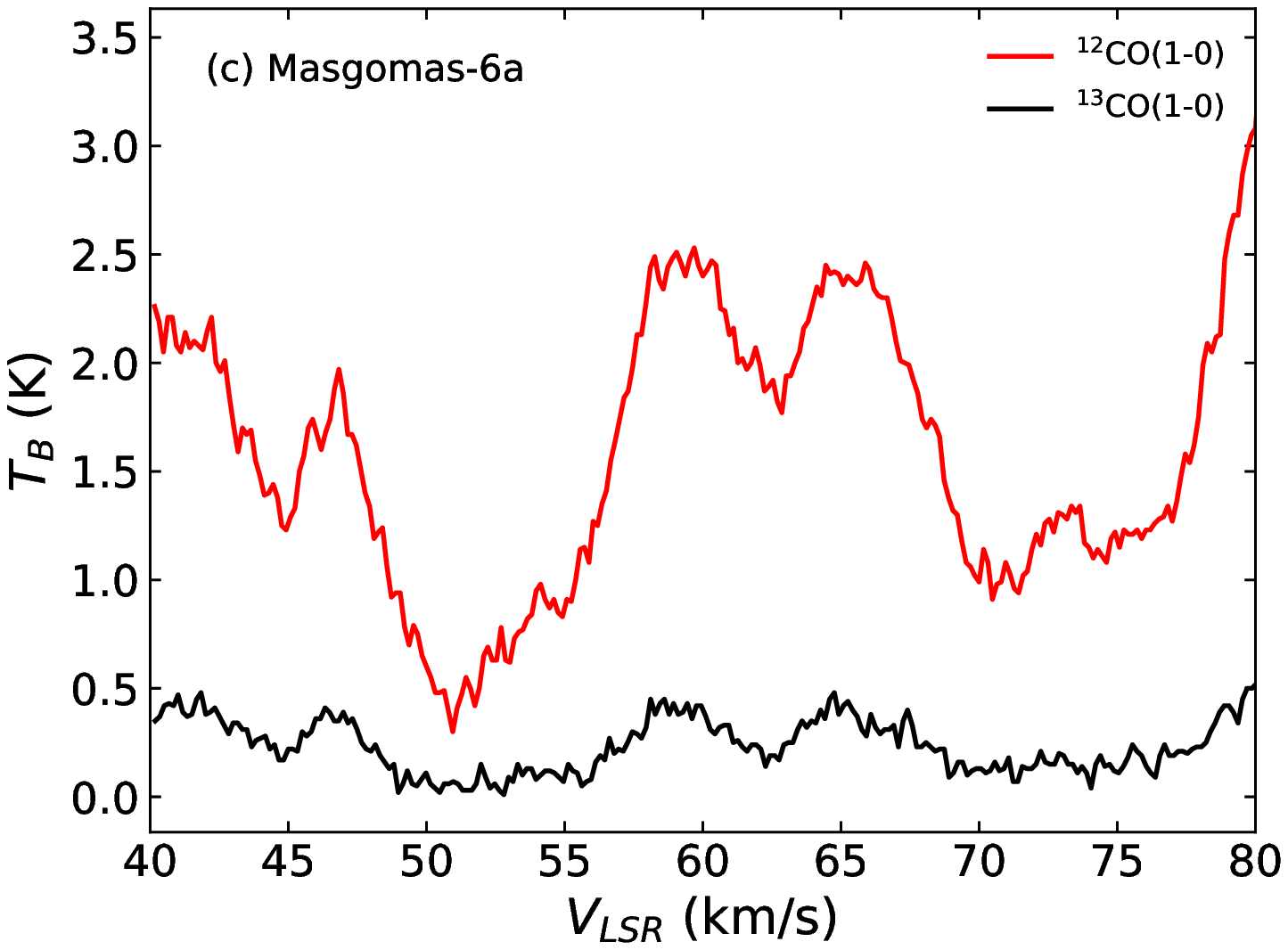}
\includegraphics[angle=0,scale=0.51]{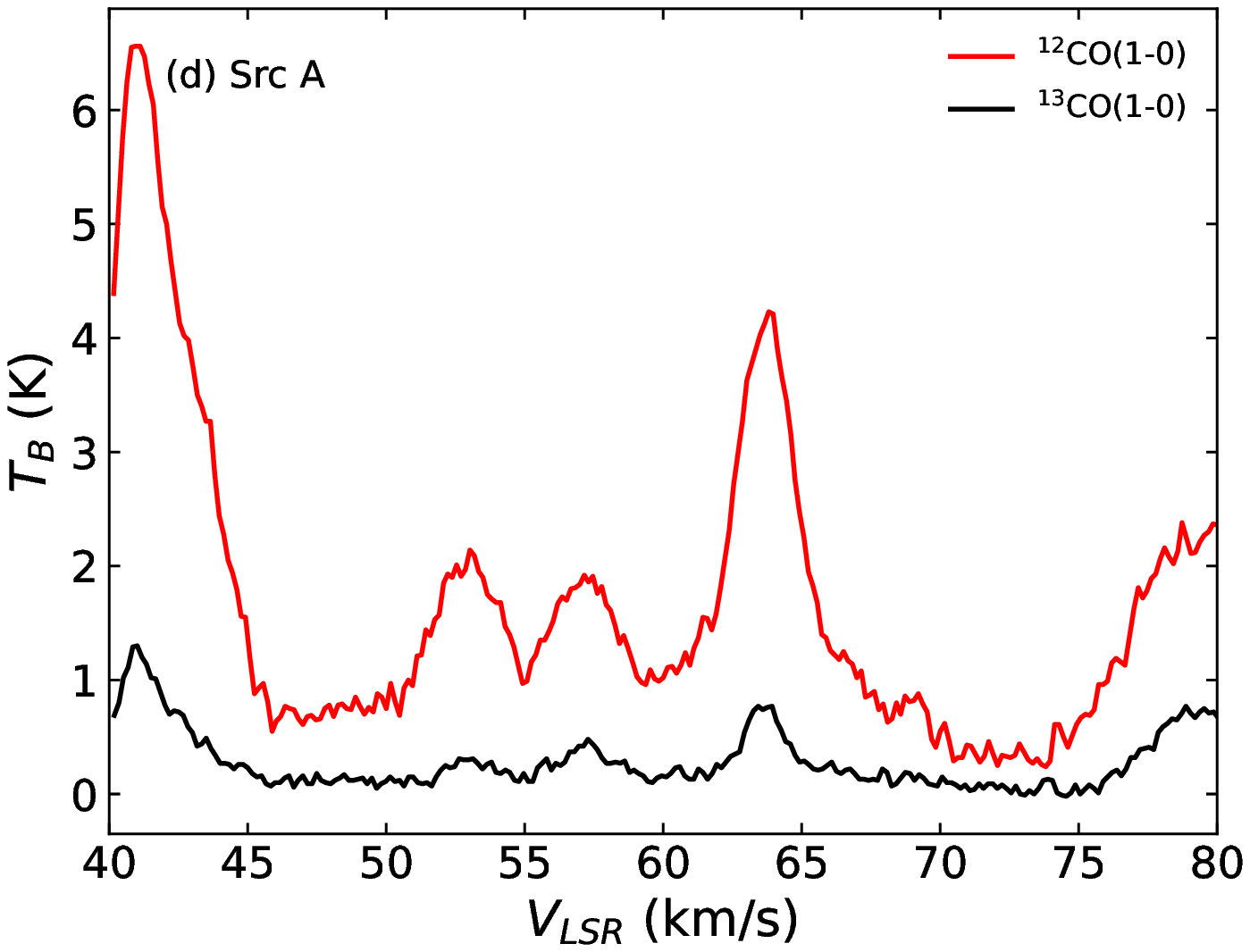}
\caption{$^{12}$CO($J$=1--0; red) and $^{13}$CO($J$=1--0; black) spectra of molecular gas toward 4FGL J1858.0+0354 (panel (a)), 4FGL J1900.4+0339 (panel (b)), Masgomas-6a (panel (c)) and SrcA (panel (d)). These spectra are extracted from regions of $0.10\degr \times 0.10\degr$ around the centre of each sources.}
\label{COsed}
\end{figure*}

The spectra of $^{12}$CO and $^{13}$CO emission toward the two 4FGL sources, Masgomas-6a and Src A are shown in Figure \ref{COsed}. We use a  Gaussian function to fit the spectra of $^{12}$CO emission around $65 \ \rm km \ s^{-1}$ toward the two 4FGL sources, Src A and Masgomas-6a. The line center and the full width at half maximum (FWHM) in the four regions are, respectively, $v_{c}=64.06 \ {\rm km \ s^{-1}}$ with FWHM $=5.17 \ {\rm km \ s^{-1}}$ for 4FGL J1858.0+0354, $v_{c}=66.03 \ {\rm km \ s^{-1}}$ with FWHM $=8.63 \ {\rm km \ s^{-1}}$ for 4FGL J1900.4+0339, $v_{c}=65.25 \ {\rm km \ s^{-1}}$ with FWHM $=6.65 \ {\rm km \ s^{-1}}$ for Masgomas-6a and $v_{c}=63.70 \ {\rm km \ s^{-1}}$ with FWHM $=2.42 \ {\rm km \ s^{-1}}$ for Src A.
Thus, one  peak at $\sim64{\rm \ km \ s^{-1}}$ is seen in the direction of  4FGL J1858.0+0354 and Src A, and another peak at $\sim 65-66{\rm \ km \ s^{-1}}$ is  seen in the direction of 4FGL J1900.4+0339 and Masgomas-6a. { This strengthens our confidence on the aforementioned correlation between the intensity of $\gamma$-ray emission and the gas distribution in the velocity range of 60 to 70 $\ \rm km \  s^{-1}$. }
In addition, We do not find any significant evidence of broadenings or asymmetries in $^{12}$CO line with respect to the narrow $^{13}$CO line, indicating no shock interaction. 

\begin{figure*}
\centering
\includegraphics[angle=0,scale=0.26]{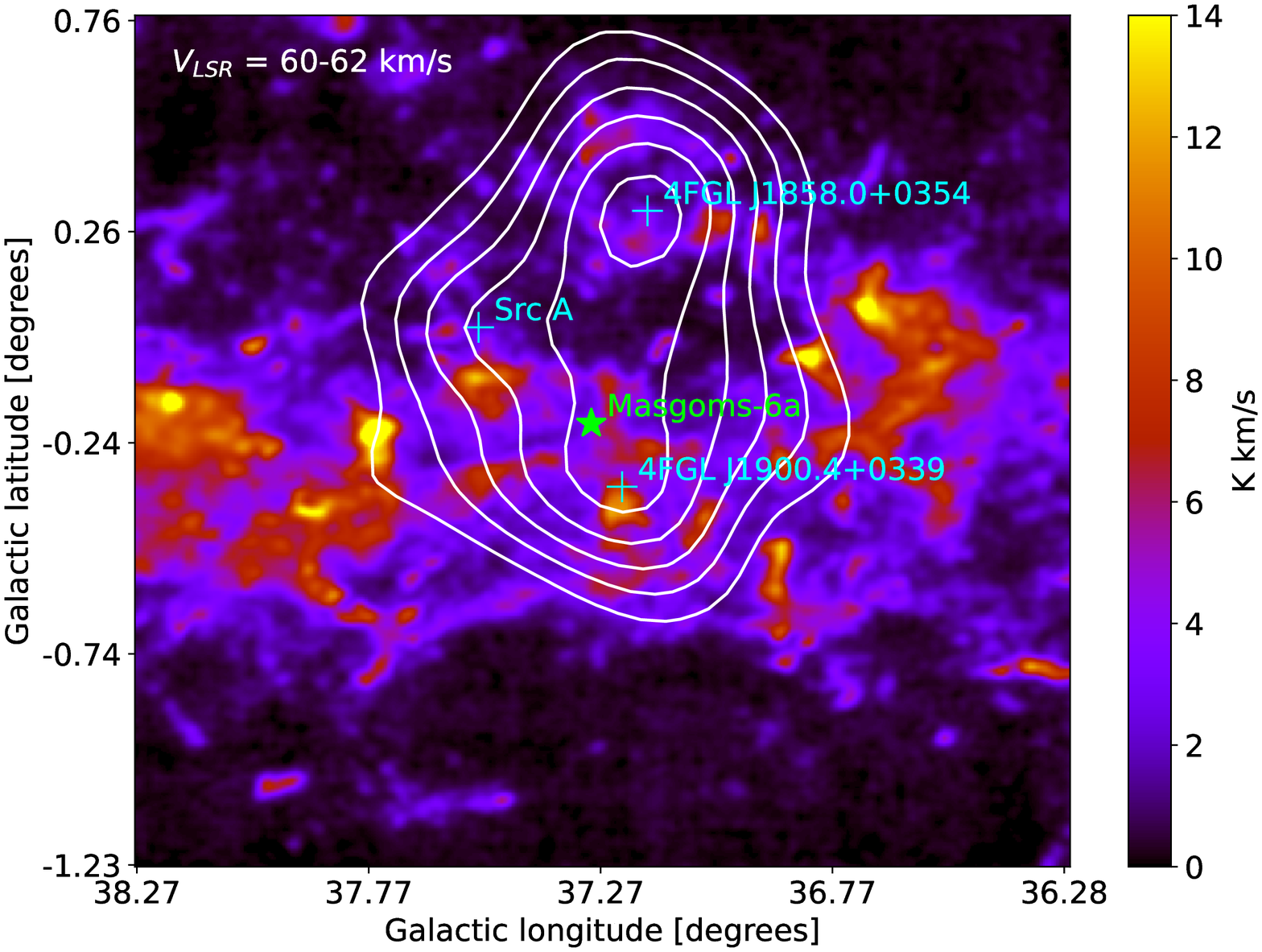}
\includegraphics[angle=0,scale=0.26]{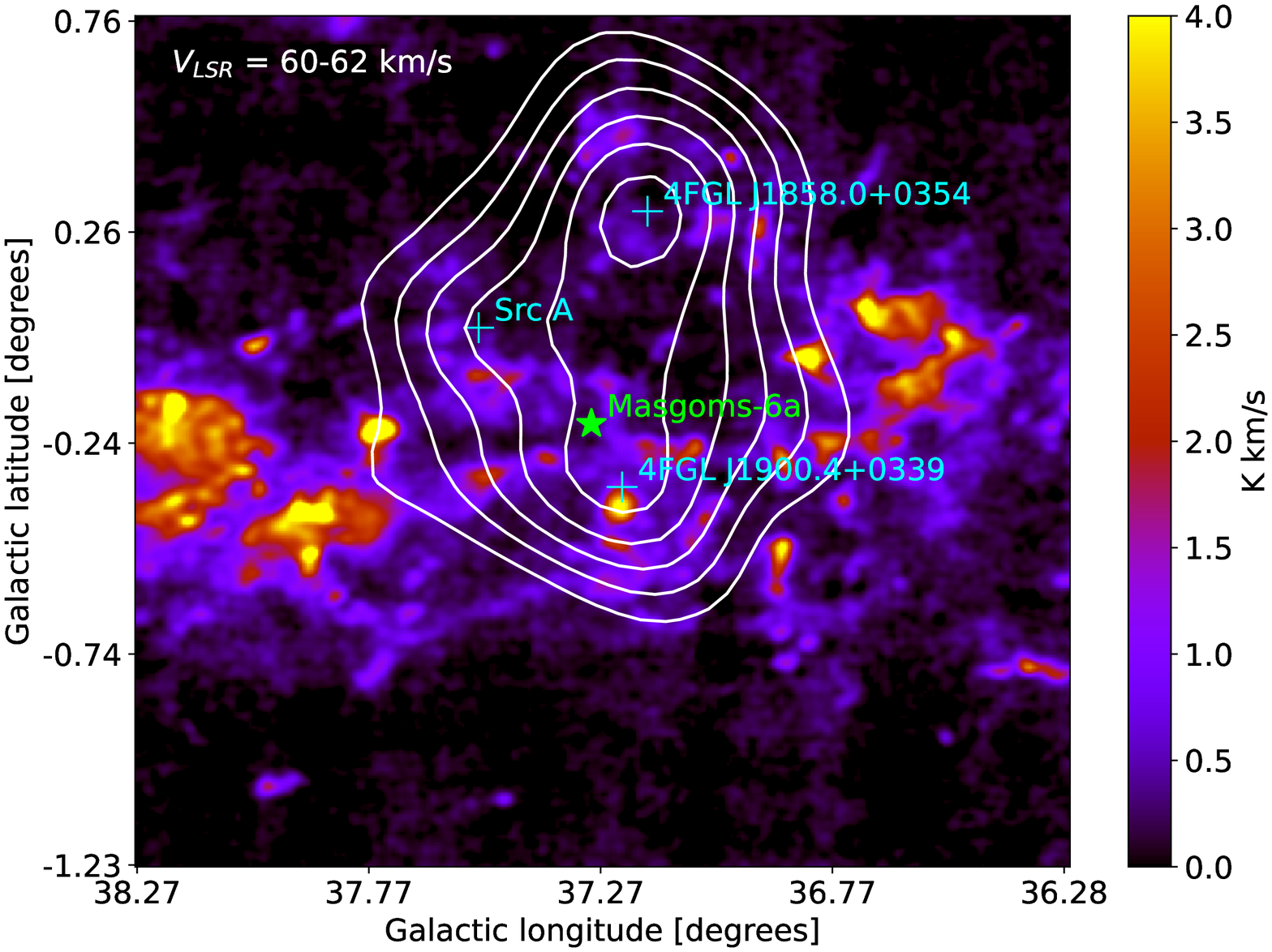}\\
\includegraphics[angle=0,scale=0.26]{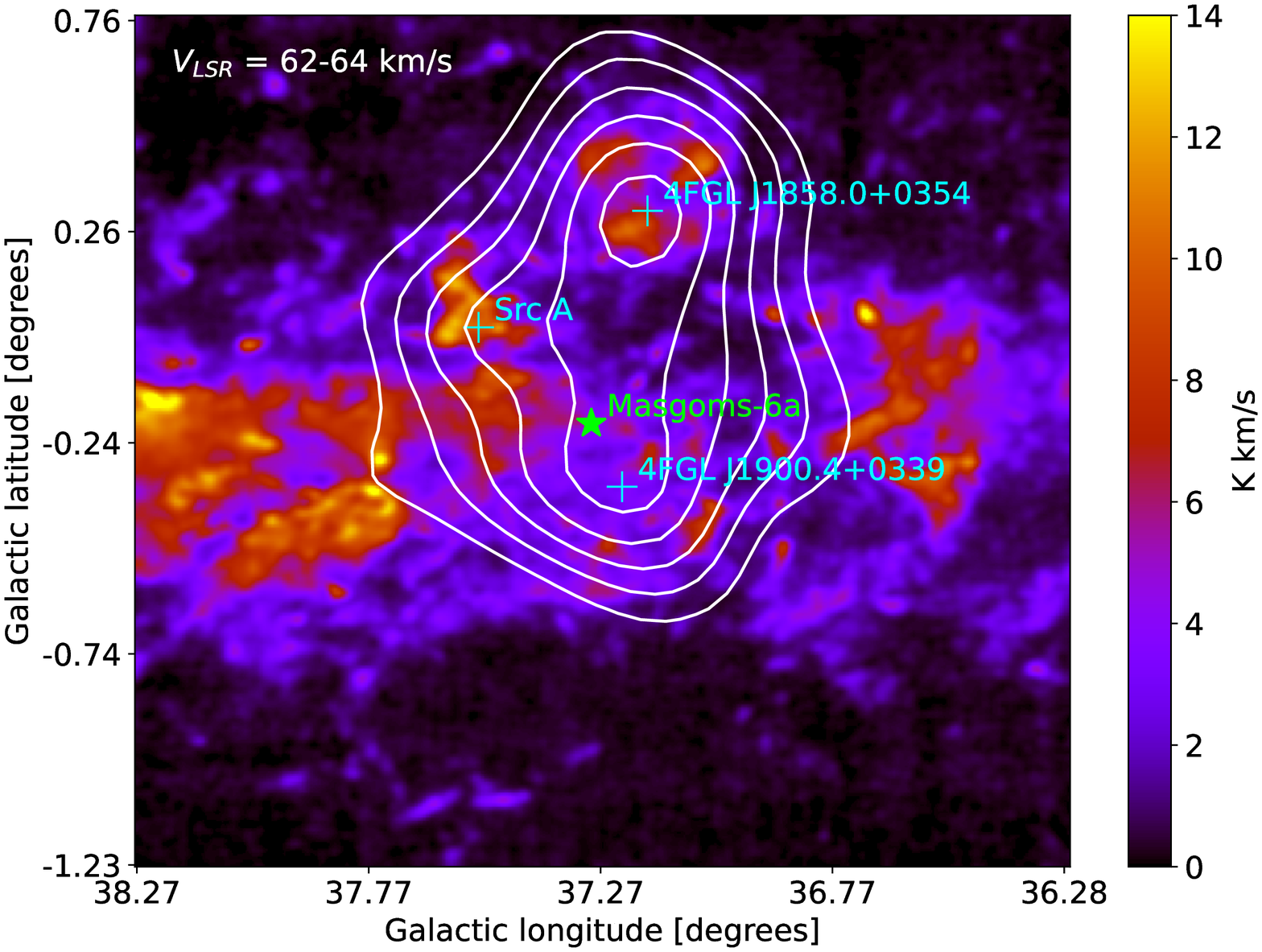}
\includegraphics[angle=0,scale=0.26]{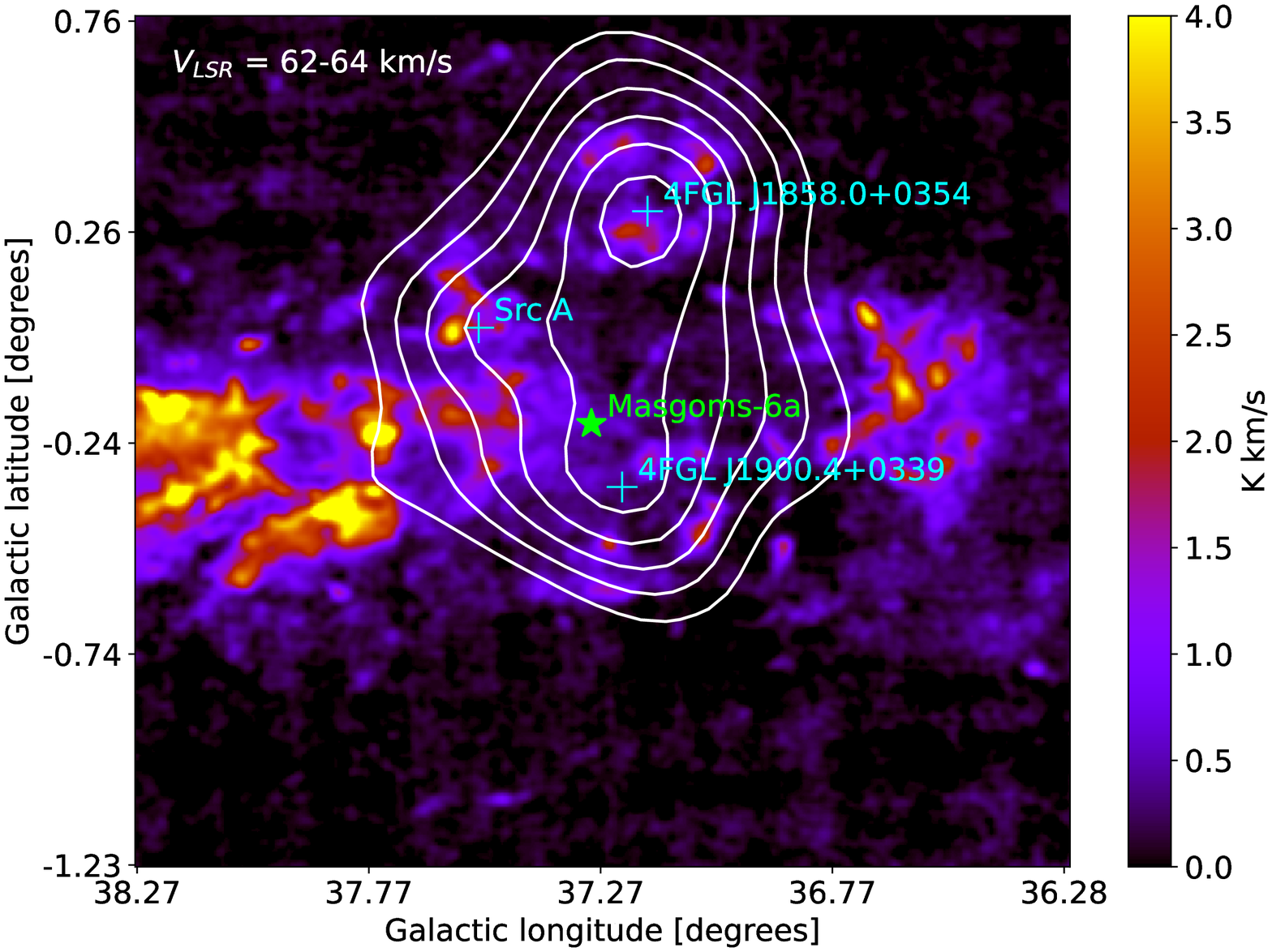}\\
\includegraphics[angle=0,scale=0.26]{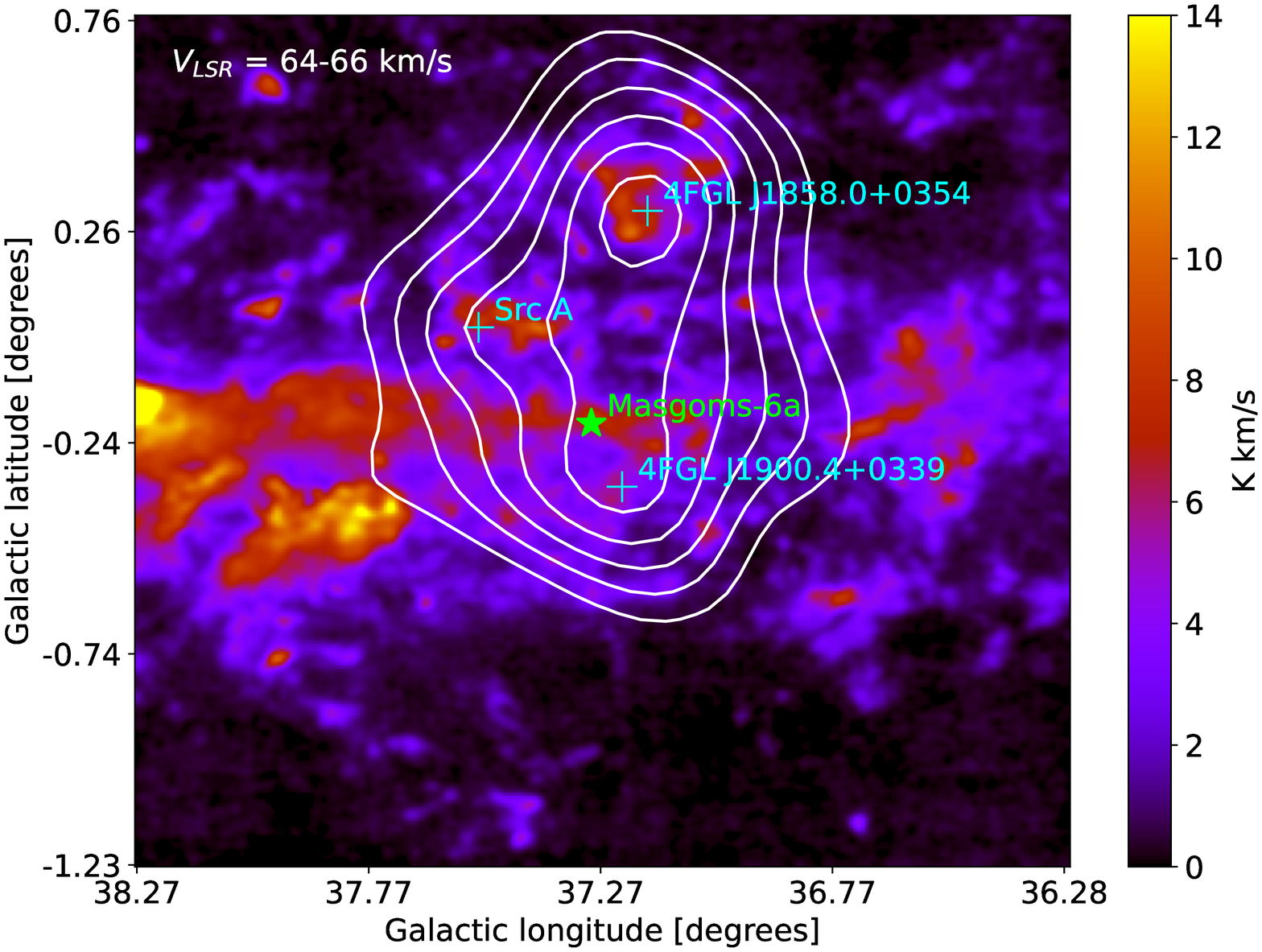}
\includegraphics[angle=0,scale=0.26]{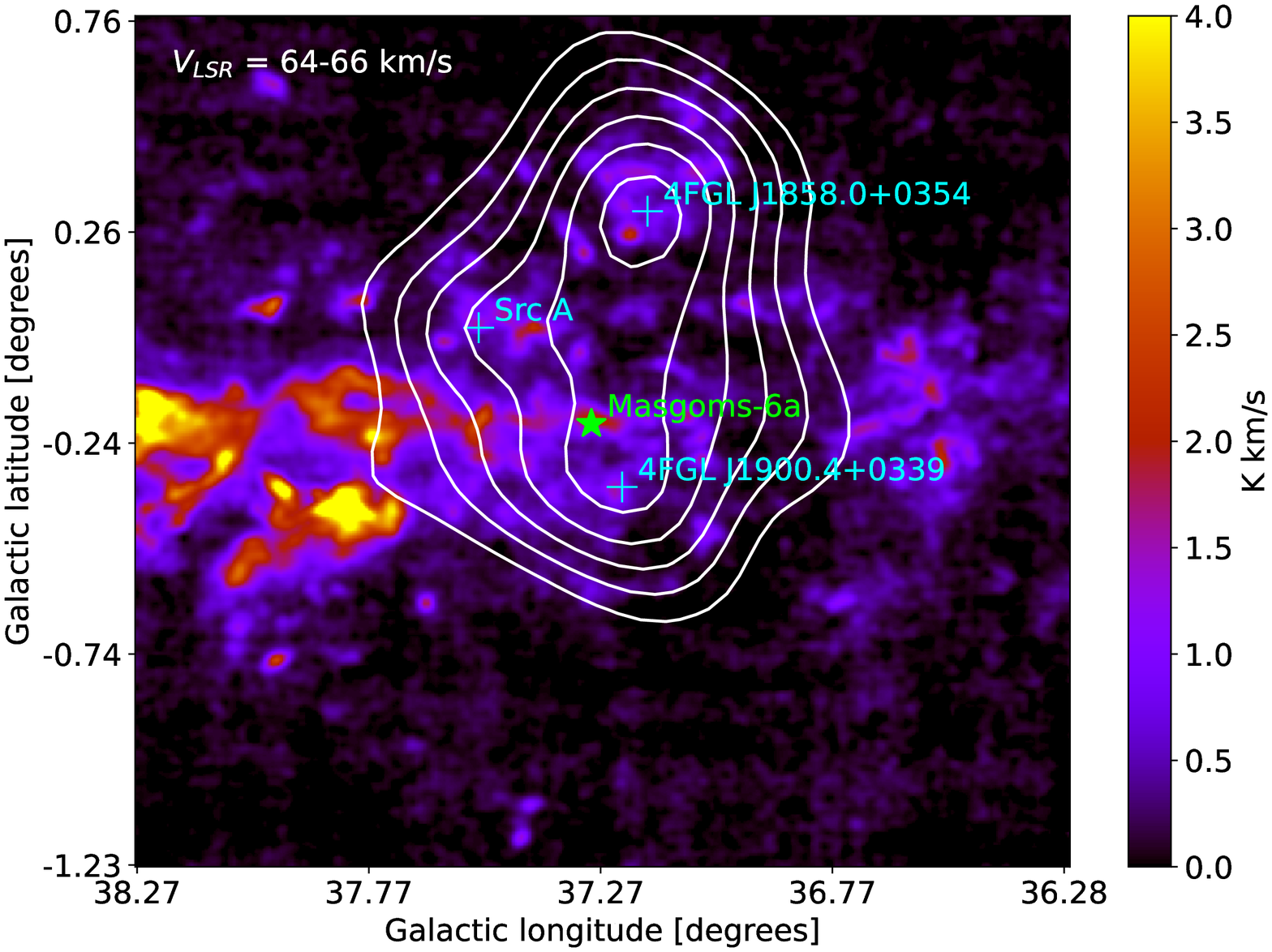}\\
\includegraphics[angle=0,scale=0.26]{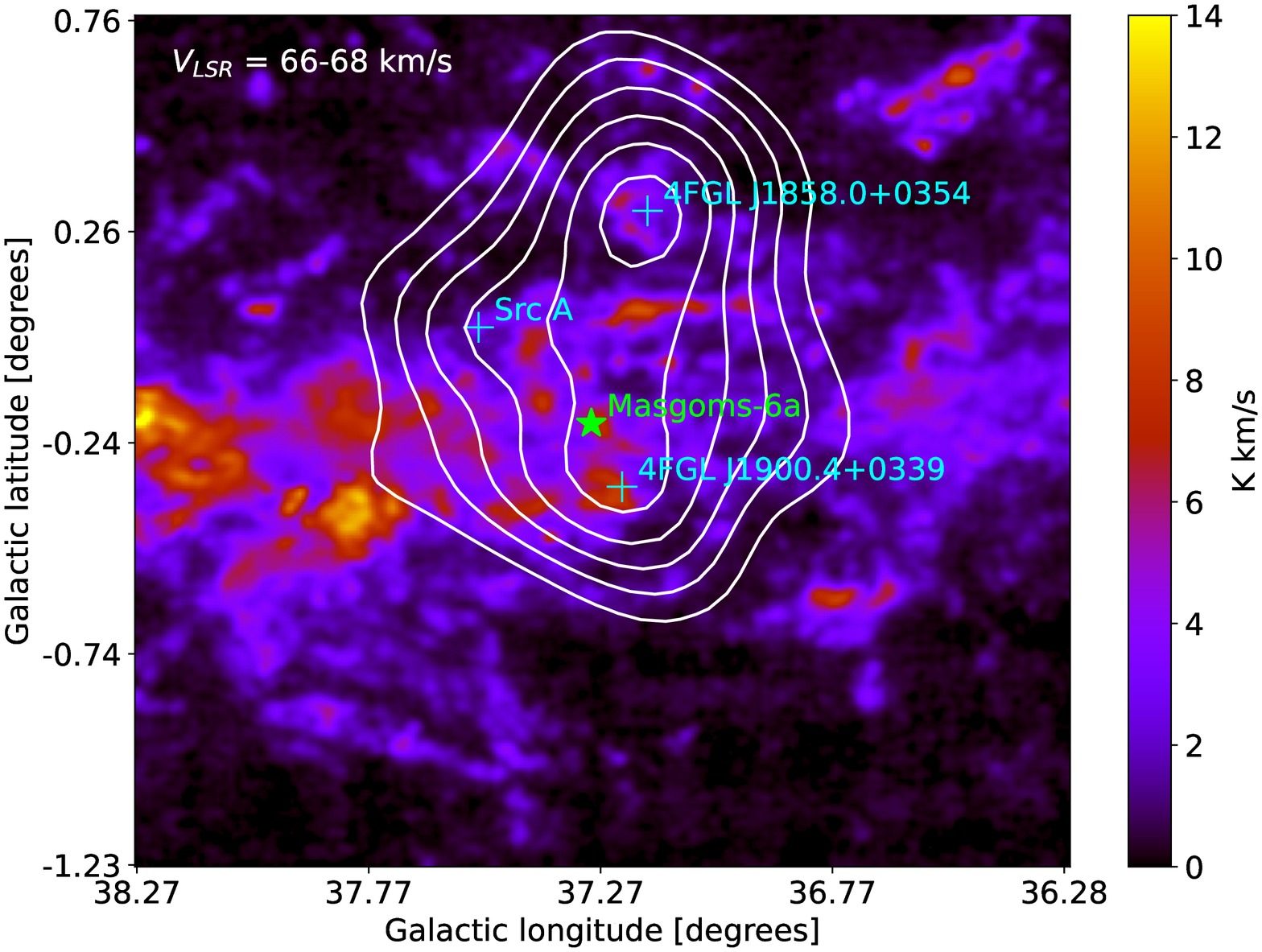}
\includegraphics[angle=0,scale=0.26]{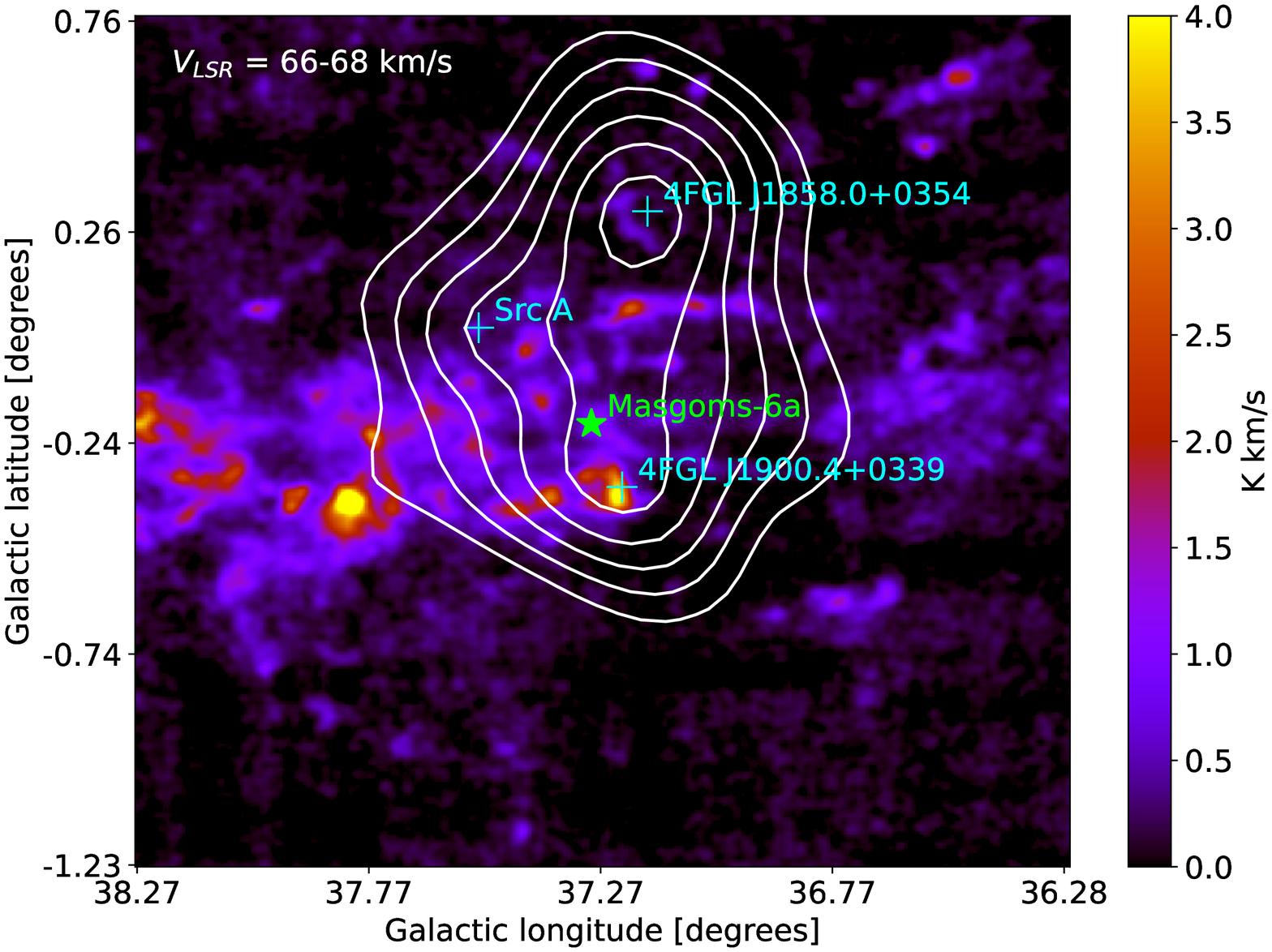}\\
\includegraphics[angle=0,scale=0.26]{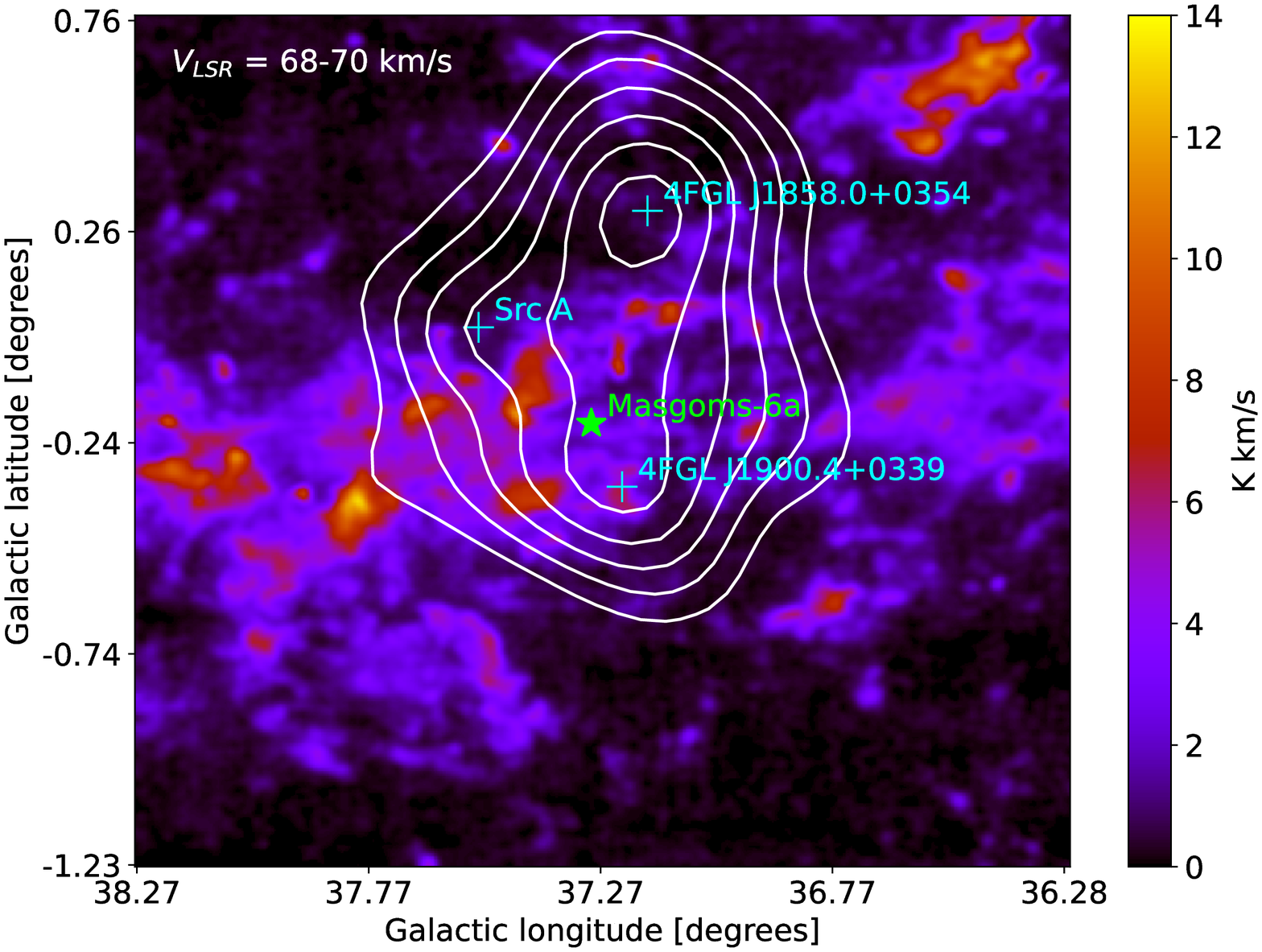}
\includegraphics[angle=0,scale=0.26]{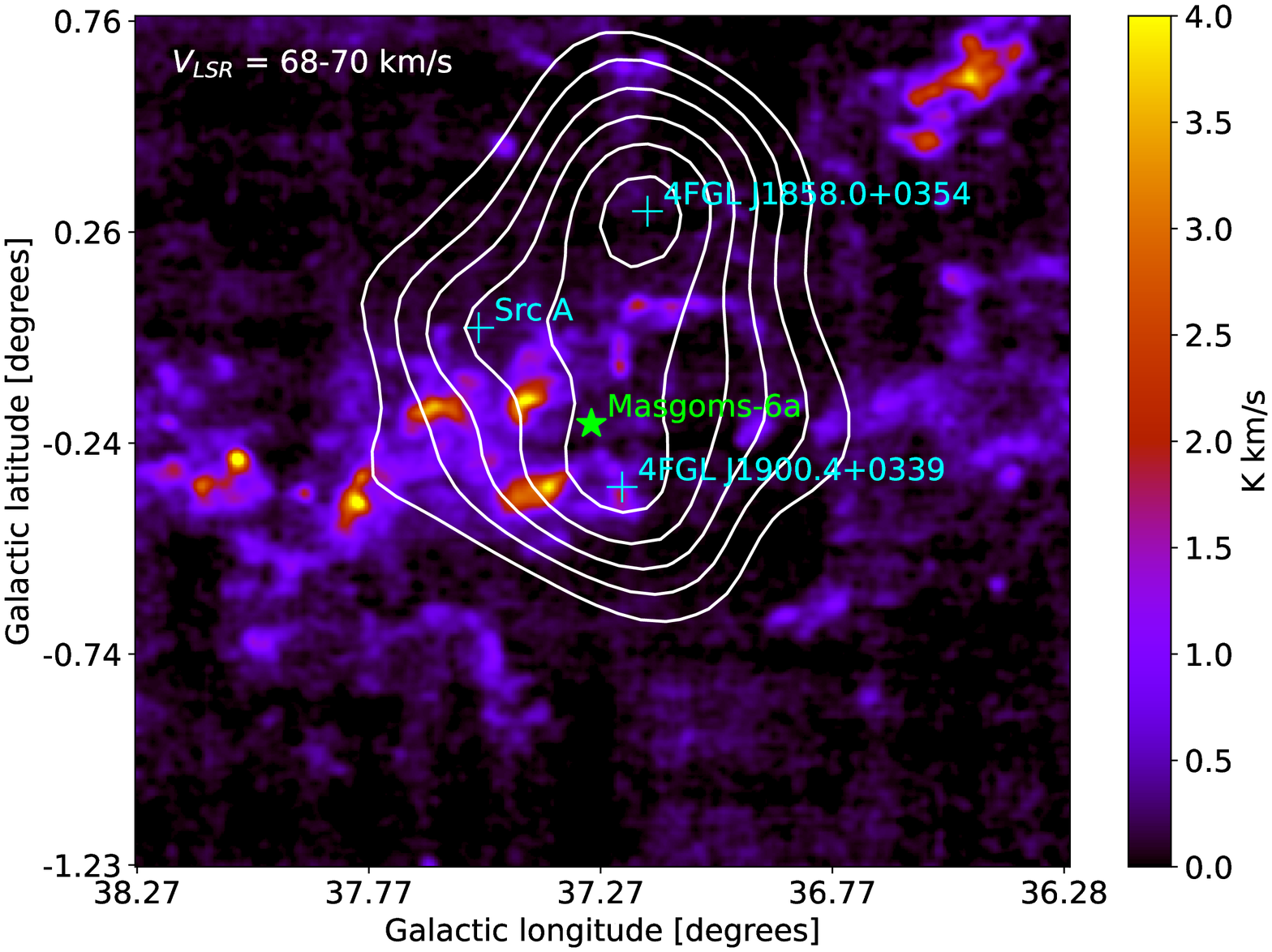}\\
\caption{$^{12}$CO($J$=1--0; left column) and $^{13}$CO($J$=1--0; right column) intensity map for five consecutive velocity ranges from 60 to 70 $\rm km \ s^{-1}$ , with a step of $2\ \rm km \ s^{-1}$ for each map. White contours correspond to {\em Fermi}-LAT significance map starting from $3\sigma$ to $8\sigma$ by $1\sigma$ steps. 
}
\label{COmap2}
\end{figure*}

Figure \ref{COmap2} displays the $^{12}$CO(J = 1-0) and $^{13}$CO(J = 1-0) maps for five consecutive velocity ranges from 60 to 70 $\rm km \ s^{-1}$ , with a step of $2\ \rm km \ s^{-1}$ for each map.
We find that the molecular gas are observed to be spatially associated with the GeV emission in all the velocity range. In particular, the correlation is the best in the velocity ranges of $62-64 \ \rm km \ s^{-1}$ and $64-66 \ \rm km \ s^{-1}$. 

Adopting the mean CO-to-$\rm H_{2}$ mass conversion factor $\rm X_{CO}=2\times10^{20} \ cm^{-2} \ K^{-1} \ km^{-1} \ s$ \citep{Bolatto2013},
we estimate that the total mass of gas within the $0.52\degr$  disk of the GeV source is about $3.13\times 10^{5}d_{3.9}^{2} \ M_{\odot}$. If we assume the { $\gamma$-ray} emission region is spherical
in geometry, then the  radius is estimated to be $R\sim36d_{3.9} \ \rm pc$.
Then the average atom density of the $\rm H_2$ cloud in this region is about $n\simeq \rm 70 d_{3.9}^{-1} \ cm^{-3}$.

\section{Interpretation of the GeV emission}

The correlation between the { $\gamma$-ray} intensity and the molecular gas density favors the hadronic origin of the { $\gamma$-ray} emission, where GeV { $\gamma$-rays} are produced by the  $pp$ interaction between the CRs and the gas. 
{We fit the { $\gamma$-ray} spectra of the source with the hadronic model using the Markov Chain Monte Carlo fitting routines of Naima, a package for the calculation 
of nonthermal emission from relativistic particles \citep{2015ICRC...34..922Z}.
First, we assume the parent proton spectrum is a single power-law  given by $dN/dE = A (E/E_0)^{-\alpha}$. 
The derived parameters are $\alpha = 2.57 ^{+0.03} _{-0.04}$, and the total energy is $W_{p}= 1.81 \times10^{49} \rm \ erg$ for the protons above $1 \rm \ GeV$, assuming the gas density is 70 $\text{cm}^{-3}$.
The best-fit result is shown by the red dashed line in Figure \ref{sed}.
Below 1 GeV, the hard spectrum is due to the $\pi^0$ bump. At higher energies ($>10 \ {\rm GeV}$), the theoretical flux exceeds the LAT data slightly (but still within the $2\sigma$ uncertainty). 
This could indicate a possible cutoff in the proton spectrum around 100 $\text{GeV}$ .
Then, we consider a cutoff power-law function for the parent proton spectrum, i.e. $dN/dE = A(E/E_0)^{-\alpha} \text{exp}(-(E/E_{\text{cut}})^\beta)$. 
The derived parameters are $\alpha=2.18 \pm 0.05$ and $W_{p}= {1.34 \times 10^{49} \rm \ erg}$ for the protons above 1 GeV, if we  take $\beta = 1$, $d=3.9 \ \rm kpc$, $n= 70 \rm \ cm^{-3} $ and $E_\text{cut} = 100 \rm \ GeV$.
The fitting result is shown by the blue line in Figure \ref{sed}.
}

The young massive star cluster candidate Masgomas-6a contains two Wolf–Rayet stars (WR122-11 and  WR122-16) and several
OB-type stars.  WR122-11 is classified as WN6 and  WR122-16 is classified as WN8-9h \citep{2018A&A...614A.116R}. These WR stars have a  mass loss rate $2-3\times 10^{-5}{\rm M_\odot yr^{-1}}$ and the wind velocity is $1000-2000 {\rm \ km \ s^{-1}}$, so the kinetic luminosity of the stellar wind is about $10^{37}{\rm \ erg \ s^{-1}}$, which
is sufficient to power the { $\gamma$-ray} emission.    As the mass-loss of WR stars are quasi-continuous over the lifetime $T$, one could expect a quasi-continuous injection of CRs into the interstellar medium over
the age of WR stars  ($T\ga 10^6{\rm yr}$). { The stellar wind from Masgomas-6a may drive a termination shock by interacting with the surrounding gas. The termination shock then accelerates cosmic rays, which produce gamma-ray emission by colliding with the molecular cloud.  }

The efficiency of production of { $\gamma$-rays} in the cloud is determined by the ratio between the diffusion timescale and the energy loss timescale of cosmic rays, i.e., $\eta_{pp}=t_{\rm diff}/t_{\rm pp}$. 
The energy loss time of cosmic ray protons colliding with the molecular gas is $t_{pp}=1/(\sigma_{\rm pp}) K_{\rm pp} n c= 10^6{\rm yr}(n/70{\rm cm^{-3}})^{-1}$, where $\sigma_{\rm pp}=3\times 10^{-26}{\rm cm^2}$ is the cross section of $pp$ collisions, $K_{\rm pp}=0.5$  is the inelasticity of pion production and $n$ is the gas number density. The diffusion time depends on the diffusion coefficient in the cloud, i.e. $t_{\rm diff}=R^2/4 D$, where $R$ is the size of the source and $D$ is the diffusion coefficient. The diffusion coefficient for CRs propagating in the molecular cloud is poorly known.  In the interstellar medium, the diffusion coefficient is of order of $10^{29}{\rm cm^2 \ s^{-1}}$
for 10–100 GeV protons
(see e.g. \citep{2007ARNPS..57..285S}) responsible for 1–10 GeV { $\gamma$-rays}. We assume a
homogeneous diffusion coefficient and parameterize it
as $D=\chi D_{\rm ISM}=\chi 10^{29}{\rm cm^2 \ s^{-1}}$ for CRs with energy of $10-100 {\rm \ GeV}$, with $\chi$
being the ratio between the CR diffusion coefficient in the cloud and the average
one of the Galactic ISM. The diffusion time is then given by $t_{\rm diff}=10^3  \chi^{-1} (D_{\rm ISM}/10^{29}{\rm cm^2 \ s^{-1}})^{-1}\,$yr for $R=36 {\rm \ pc}$. Thus, the required CR injection rate to power the { $\gamma$-ray} emission is $L_p=L_\gamma/\eta_{pp}=1.8\times 10^{38}\chi {\rm \,erg \ s^{-1}}$. To account for the CR injection rate with the stellar wind power of WRs, $\chi\la 0.1$ is required for a reasonable CR acceleration efficiency, implying that the diffusion coefficient is significantly suppressed in the molecular cloud. We also note that  a significant part of CRs could have already escaped out of the { $\gamma$-ray} emission region. The relation between the total injected CR energy and the CR energy remained in the { $\gamma$-ray} emission region can be expressed as
$L_{\rm {p,tot}}/L_p=(r_{\rm diff}/R)^2=100 (\chi/0.1) (T/10^6{\rm yr})$, where $r_{\rm diff}$ is the  the total diffusion length of CRs corresponding to the age $T$ of the WR stars. 

Supernova remnants (SNRs) could also be a potential accelerator of cosmic rays in this region, since the energy release in CRs from a single SNR is $10^{49}-10^{50}{\rm erg}$. The CRs escaping from SNRs can produce { $\gamma$-rays} by
interacting with the surrounding gas\citep{Aharonian1996,Rodriguez Marrero2008,Aharonian2008a,Gabici2009,Uchiyama2012,Zhang2021}. No recorded SNRs is found in this region and future multi-wavelength observations may be helpful to determine this scenario. The radial distribution of { $\gamma$-rays}  can provide key information for the injection history of CRs, which is useful to distinguish between the continuous stellar wind scenario and the single SNR scenario \citep{2019NatAs...3..561A}. {Pulsar wind nebulae (PWN) are another potential sources for the { $\gamma$-ray} emission. There is a middle-aged pulsar PSR J1858+0346 located positionally close to the center of the { $\gamma$-ray} sources, as marked by the grey dot in Figure \ref{GeVmap0}, with a characteristic age of 2\,Myr and a distance of 5.5\,kpc \citep{2005AJ....129.1993M}. If the { $\gamma$-ray} source is related to the PWN of the pulsar, the { $\gamma$-ray} luminosity would be about $3.5\times 10^{35}\,\rm erg/s$, and cannot be explained based on the current spin-down power of PSR J1858+0346, which is $4.8\times10^{33}\,\rm erg/s$. However, we note that the cooling timescale of GeV-emitting electrons, $t_c\approx 5(E_e/100{\rm GeV})^{-1}(B/5\mu\rm G)^{-2}\,$Myr, is comparable to or even longer than the age of the pulsar assuming a typical interstellar magnetic field strength. As a result, considering a much higher spin-down power of the pulsar at its early age, we cannot exclude the pulsar/PWN origin of the { $\gamma$-ray} source.}

\section{Summary}

\begin{figure*}
\centering
\includegraphics[angle=0,scale=0.7]{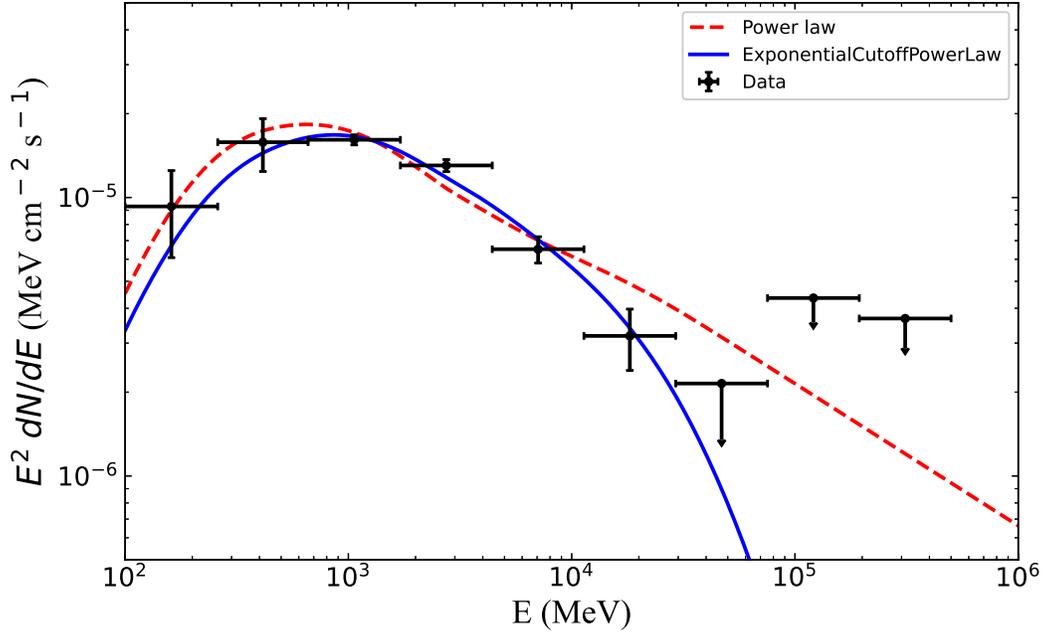}
\caption{Modeling of the SED of the { $\gamma$-ray} source with a  hadronic model. { The black dots represent the \emph{Fermi}-LAT data in 0.1-500 GeV.} The red dash line represents the modeling assuming a single power-law proton spectrum. 
Model parameters are $\alpha = 2.57 ^{+0.03}_{-0.04}$ (the proton spectral index) and $W_{p}= 1.81 \times10^{49} \rm \ erg$ (the total energy in protons above 1 GeV).
The blue solid line represents the modelling assuming a  power-law with a high-energy cutoff for the proton spectrum.
Model parameters are $\alpha=2.18 \pm 0.05$, $\beta = 1$,  $E_\text{cut} = 100 \rm \ GeV$, and $W_{p}= {1.34 \times 10^{49} \rm \ erg}$ for the protons above 1 GeV. The distance of the source is taken to be $d=3.9 \ \rm kpc$ and the gas number density is taken to be $n= 70 \rm \ cm^{-3} $.
}
\label{sed}
\end{figure*}

Massive star clusters hosting WR and OB-type stars have been suggested to be accelerators of Galactic CRs. 
We analyzed the \emph{Fermi} Large Area Telescope's data towards the direction of Masgomas-6a, a young massive star cluster candidate hosting two WR stars, which is located at a distance of $3.9{\ \rm kpc}$. An extended { $\gamma$-ray} source with a radius of $0.52$ degree is found in the vicinity of Masgomas-6a. { Our analysis shows a correlation between the intensity of $\gamma$-ray emission and the density of molecular gas in the region.} Combined with the fact that the distance of the molecular cloud inferred from the velocity distribution is consistent with that of Masgomas-6a, the correlation indicates that the { $\gamma$-ray} emission is related with Masgomas-6a. A natural scenario is that the two WR stars in Masgomas-6a accelerate cosmic rays, which produce the { $\gamma$-ray} emission through inelastic proton-proton collisions with the molecular gas. With a kinetic luminosity of $\sim 10^{37}{\rm erg \ s^{-1}}$ in the stellar winds,  the two WR stars are capable of powering the { $\gamma$-ray} emission. The radius of the GeV source is about $36 {\rm \ pc}$ for a source distance of $3.9{\ \rm kpc}$. Assuming a normal CR diffusion coefficient, the vast majority of CRs accelerated by WR stars in their lifetime must have escaped out of the source as the injection of CRs  is expected to last million years. The energy budget constraint  suggests a CR diffusion coefficient smaller than that in the Galactic ISM, indicating a strong suppression of CR diffusion when they are propagating in the  molecular cloud. { The spectral data shown in Figure \ref{sed} seems to suggest  the  presence of a cutoff in the proton spectrum  around 100 GeV, but we note that the errors of the data at high energies are rather large, so one can not exclude a single power-law  spectrum. Thus, the maximum energy of cosmic rays accelerated by Masgomas-6a should be least  {$\sim 100$}{\rm \  GeV}.}

\acknowledgments
{ We would like to thank the referee for the constructive report.}
This work was supported by the National KeyR \& D program of China under the grant 2018YFA0404203, the NSFC grants 12121003 and U2031105, and China Manned Spaced Project
(CMS-CSST-2021-B11).
This research made use of the data from the Milky Way Imaging Scroll Painting (MWISP) project, which is a multi-line survey in $^{12}$CO/$^{13}$CO/C$^{18}$O along the northern galactic plane with PMO-13.7m telescope. { Thanks to NASA Fermi Collaboration for providing the excellent data and tools that made this work possible.} We are grateful to all the members of the MWISP working group, particularly the staff members at PMO-13.7m telescope, for their long-term support. MWISP was sponsored by National Key R\&D Program of China with grant 2017YFA0402701 and CAS Key Research Program of Frontier Sciences with grant QYZDJ-SSW-SLH047.

\begin{table}[ht!]
\caption{Morphological models tested for the GeV { $\gamma$-ray} emission above 3 GeV.}
\begin{center}
    \begin{tabular}{lccccccc}
        \hline
        Morphology ($>$ 3 GeV) &R.A.(degrees) & Decl.(degrees) & Extension & loglikelihood & TS & $\rm TS_{ext}$ & $N_{\rm dof}$\\ \hline
       Null           & --  & --& --  & 119936.52   & -- & -- & -- \\
        PS+PS           & --  & --& --    & 120016.56 & 160.08 & -- & 8 \\
         Disk    & $284.81\degr \pm 0.08\degr$ & $3.84\degr \pm 0.04\degr$ &   $0.43\degr_{-0.03\degr}^{+0.02\degr}$   & 120028.15	  &  183.26 &  129.80 &5\\
        Gaussian   & $284.82\degr \pm 0.04\degr$    & $3.84\degr \pm 0.04\degr$ & $0.46\degr \pm 0.04\degr$   &120025.90       & 178.76 & 125.30 &5\\
         \hline
    \end{tabular}
    \end{center}
    \tablenotetext{}{{ Note:} PS, Disk and Gaussian represent, respectively, a point-like source model, an uniform disk model and a two-dimensional Gaussian model.  PS+PS  represents the two point source model with source positions at the two 4FGL sources. 
    Extension in the disk and Gaussian models represents the radius containing 68\% of the intensity of the tested models. $N_{\rm dof}$ represents the number of degrees of freedom for each model.}
    \label{tab:morph}
\end{table}{}

\clearpage

\end{document}